\documentclass[12pt,preprint]{aastex}
\begin{document}
\title{The Extreme Ultraviolet Variability of Quasars}
\author{Brian Punsly\altaffilmark{1}, Paola Marziani\altaffilmark{2}, Shaohua Zhang \altaffilmark{3}, Sowgat Muzahid \altaffilmark{4} and Christopher P. O'Dea\altaffilmark{5,6}}\altaffiltext{1}{1415 Granvia Altamira, Palos Verdes Estates CA,
USA 90274 and ICRANet, Piazza della Repubblica 10 Pescara 65100,
Italy, brian.punsly@cox.net} \altaffiltext{2}{INAF, Osservatorio
Astronomico di Padova, Italia}\altaffiltext{3}{Antarctic Astronomy
Division Polar Research Institute of China 451, Jinqiao Rd.,
Shanghai, China} \altaffiltext{4}{The Pennsylvania State University,
State College, PA 16802 , USA}\altaffiltext{5}{Department of Physics
and Astronomy, University of Manitoba, Winnipeg, MB R3T 2N2
Canada}\altaffiltext{6}{School of Physics \& Astronomy, Rochester
Institute of Technology, Rochester, NY 14623, USA}
\begin{abstract}
We study the extreme ultraviolet (EUV) variability (rest frame
wavelengths 500 - 920 $\AA$) of high luminosity quasars using HST
(low to intermediate redshift sample) and SDSS (high redshift
sample) archives. The combined HST and SDSS data indicates a much
more pronounced variability when the sampling time between
observations in the quasar rest frame is $> 2\times 10^{7}$ sec
compared to $< 1.5\times 10^{7}$ sec. Based on an excess variance
analysis, for time intervals $< 2\times 10^{7}$ sec in the quasar
rest frame, $10\%$ of the quasars (4/40) show evidence of EUV
variability. Similarly, for time intervals $>2\times 10^{7}$ sec in
the quasar rest frame, $55\%$ of the quasars (21/38) show evidence
of EUV variability. The propensity for variability does not show any
statistically significant change between $2.5\times 10^{7}$ sec and
$3.16\times 10^{7}$ sec (1 yr). The temporal behavior is one of a
threshold time interval for significant variability as opposed to a
gradual increase on these time scales. A threshold time scale can
indicate a characteristic spatial dimension of the EUV region. We
explore this concept in the context of the slim disk models of
accretion. We find that for rapidly spinning black holes, the radial
infall time to the plunge region of the optically thin surface layer
of the slim disk that is responsible for the preponderance of the
EUV flux emission (primarily within 0 - 7 black hole radii from the
inner edge of the disk) is consistent with the empirically
determined variability time scale.

\end{abstract}

\section{Introduction}
The spectral energy distribution (SED) of quasars typically peaks
around 1100\AA\ then decays in the extreme ultraviolet (EUV)
shortward of this peak \citep{tel02,ste14}. The broadband optical/UV
emission is widely believed to be optically thick emission from gas
accreting onto a central supermassive black hole \citep{sun89}. The
emission naturally arises in an inflow from the differential
shearing of the infalling gas \citep{lyn71}. For orbital velocities
on the order of the local Keplerian angular velocity, the angular
velocity increases towards the black hole with an ever increasing
gradient. Thus, the maximum dissipation and the highest frequency
thermal emission arises from the innermost regions of the accretion
flow \citep{sha73}. The EUV emission just shortward of the SED peak
(rest frame wavelengths 500 - 1000 $\AA$) should be predominantly
from the innermost optically thick accretion flow
\citep{sun89,szu96,zhu12}. In this paper, we study the time
variability behavior of the EUV continuum emission. We also
speculate on the size of the EUV emitting region based on slim
accretion disk models and our temporal analysis.

\par The EUV continuum is a uniquely important region of the quasar spectrum for two reasons. Firstly, an estimate of the size of the EUV region can be
compared to the dimension expected from accretion disk theory in
order to critically analyze the qualitative applicability of
accretion disk theory to quasar accretion flows near the central
supermassive black hole. In particular, as noted above, being the
highest energy optically thick emission should place this region
near the inner edge of the accretion disk. If accretion disk models
are appropriate then an estimate of the location of the EUV emitting
region should indicate that the EUV emitting region lies close to
the inner edge of the disk. The EUV region is a modest fraction of
the total optically thick accretion flow luminosity, $\sim 5\% -
10\%$ \citep{pun14}. This provides a second constraint on accretion
disk models. The small total EUV luminosity seems to favor a
dimension of the EUV emitting region comparable to those of the
inner edge of the disk as opposed to a dimension that is an order of
magnitude or more larger than the inner edge of the disk
\citep{sun89,zhu12}. We will critically evaluate our size estimates
of the EUV region derived in this article in the context of slim
accretion disk theory. The second unique property of the EUV
continuum is a consequence of its connection to the launching of
powerful relativistic jets in quasars. There exists a strong
correlation between jet power and EUV spectral index that has been
recently demonstrated in radio loud quasars
\citep{pun14,pun15,pun16}. In particular, the long term time average
jet power, $\overline{Q}$, is correlated with the spectral index in
the EUV, $\alpha_{EUV}$; defined in terms of the flux density by
$F_{\nu} \sim \nu^{-\alpha_{EUV}}$ computed between 700\AA\, and
1100\AA\,. Larger $\overline{Q}$ tends to decrease the EUV emission.
In general, lobe dominated radio loud quasars tend to have larger
values of $\alpha_{EUV}$ than radio quiet quasars. The
straightforward implication is that the EUV emitting region is
related to the jet launching region in quasars. Therefore, an
estimate of the size and location of the optically thick EUV
emitting region yields a constraint on the size and location of the
jet launching region. These two aspects indicate the potential power
of estimating the size of the EUV emitting region.

\par The dimensions of the emitting for the EUV continuum region is many orders of magnitude
smaller than the highest spatial resolution of optical telescopes.
Consequently, observations cannot directly constrain its size.
However, time variability can be used as a crude probe of the size
of the region. We use multi-epoch archival UV observations with the
Hubble Space Telescope (HST) of quasars with a redshift $0.4<z<2.0$
and Sloan Digital sky Survey (SDSS) of high redshift quasars $3.3 <
z < 4.0$ in order to find the EUV variability time scales in the
quasar rest frame. Our sample is restricted to time intervals of
less than one year in the quasar rest frame. The combined data-set
of HST and SDSS observations presented here shows that a dramatic
change in the behavior of variability occurs on times scales less
than one year and the sole goal of this effort is to achieve the
maximum temporal resolution of this region. As such, we do not
investigate duty cycles (i.e., multiple variations over a few years
that can approximately cancel out the total variation) or power
spectra of the variability. It is not that such studies are not
interesting in their own right, but our analysis yields a simple
clean result independent of the interpretive filter of any analysis
tool. Our sample of data is small for the purposes of a time series
analysis and does not possess a uniform sampling of time intervals.
Power spectra are often difficult to interpret when based on small
number statistics and even more so with sparse and irregular
temporal sampling \citep{kit61,nei96}. Consequently there is not
much motivation to study time intervals longer than one year in the
quasar rest frame for this study.
\par Previous studies of quasar variability have focused on the UV and optical bands in the
quasar rest frame. In \citet{wil05} composite spectra from $1000\AA$
- $6000\AA$ in the low and high phases were compared for a sample of
315 highly variable SDSS quasars. The study showed that the blue end
of the spectra were more variable. However, the only variability
data as a function of time for the entire SDSS sample of $\sim 2500$
objects was based on a broad band flux integration. Because of the
broad line contamination, this type of analysis does not extract the
time dependent variability of the continuum - our primary goal in
the EUV. SDSS photometry and Palomar Sky Survey photometry were
compared in order to study the time dependent variability in quasars
in the UV and optical bands \citep{mac12}. Broad band photometry was
also implemented in \citet{wel11} using GALEX data in order to
explore far UV and near UV variability in quasars. Broad band
photometry suffers from broad line contamination so one does not
study the continuum variability explicitly. There was an
International Ultraviolet Explorer variability study of 21 quasars
based on broadband photometry (diClemente 1995). This was typically
far UV and near UV data, but $\sim 5$ objects had spectra that
covered the EUV. Once these noisy spectra were integrated over
wavelength there was large contamination from broad emission lines
so this small study is not of much value for our purposes. Thus,
this is the first study of the variability of the continuum in
quasars on the high frequency side of the peak of the spectral
energy distribution.
\par In Section 2,  we describe the HST observations with a
particular emphasis on the criteria for an accurate flux
calibration. Section 3 is a similar analysis of the SDSS archives.
The next section combines the two samples and analyzes the data
scatter. In Section 5, we compare our results to slim disk accretion
flow models.

\section{HST Observations} Hubble Space Telescope (HST) observations
can reliably detect the quasar EUV continuum below the Lyman
continuum if $z>0.35$. In principle, HST observations can be used to
monitor the EUV continuum variability of quasars over the last
quarter century. There are two issues that limit the use of the HST
for EUV monitoring of quasars. Firstly, the long exposures necessary
with HST for individual quasar UV spectra make the time allocations
required for long term UV monitoring prohibitively large. Thus,
monitoring of the quasar UV spectra consumes too much of the the
available resource and is not a viable proposal topic. Therefore,
variability can only be found serendipitously. Secondly, in order to
detect variability, reliable flux calibration is required. This is a
major concern for narrow apertures in general and wider apertures if
the source is not accurately aligned with the slit center.
\par There were many quasar observations with the Faint Object Spectrograph (FOS).
However, the alignment issue was particularly severe in FOS
pre-COSTAR observations (see FOS Instrument Science Reports
CAL/FOS-107 and CAL/FOS-122). The two main target acquisition
methods used for FOS are ACQ/BINARY and ACQ/PEAK. Often, in order to
save observing time, the target acquisition sequence only included
ACQ/BINARY which tries to locate the object based on the input
coordinates. However, the pre-COSTAR alignment was not adequate for
such an expedience. Thus, many of the observations are poorly
centered and seemed to indicate variability of the quasar, but in
fact one is only seeing the results of instrumental error. The most
reliable target acquisition sequence involves ACQ/BINARY followed by
ACQ/PEAK. The routine ACQ/PEAK is an iterative process designed to
position the target at the center position of the brightest grid
point of the detector diode array. The nominal centering error for
ACQ/BINARY alone (not followed by ACQ/PEAK) is $\approx 0.12"$ and
the nominal centering uncertainty for a ACQ/BINARY followed by
ACQ/PEAK acquisition is 0.05" \citep{eva04}. If ACQ/BINARY was
followed by ACQ/PEAK, this acquisition sequence has another smaller
centering issue. ACQ/PEAK was generally not performed with the same
aperture that the measurement was made in and for pre-COSTAR
observations these misalignments were not well known. Due to the
alignment uncertainties, Pre-COSTAR, the smallest aperture that is
likely to yield a reliable flux measurement is 1.0" if a ACQ/BINARY
followed by ACQ/PEAK acquisition occurs. If the the acquisition
sequence is simply ACQ/BINARY then the minimum aperture is 4.3" for
a reliable flux measurement. Pre-COSTAR observations in the 4.3"
aperture flux measurements will typically be reliable to $< 10\%$
level from the point of view of aperture correction errors due to
target mis-centering. For smaller apertures the centering
uncertainty combined with the uncertainty in aperture alignments
produces an absolute flux uncertainty that is too large for their
consideration in this treatment if there is no ACQ/PEAK step in the
acquisition sequence.  If the acquisition and the observation were
both in the same aperture some of this uncertainty is removed for
smaller apertures (such as 1.0"). However, this instance never
occurred in the observations that we considered. We looked at the
acquisition files for flags that the flux measurement was suspect.
Sometimes there are two spectra taken with two different gratings
during the same observation that overlap in frequency. If the flux
mismatch between the gratings in this overlap is less than 10\% this
is suggestive of a reliable flux calibration even for an observation
with no ACQ/PEAK. We excluded observations with the 0.25" x 2.0"
slit in all instances because of the impact of poor target
centering. The efficiency of the detectors is very low at the
extreme blue end of the G130H grating coverage, and there can be
significant shot noise that distorts the true continuum at the
shortest wavelengths (Ian Evans private communication 2015). Thus,
the blue end of the G130H grating was not considered suitable for
accurate absolute flux measurements.
\begin{figure}
\begin{center}
\includegraphics[width=125 mm, angle= 0]{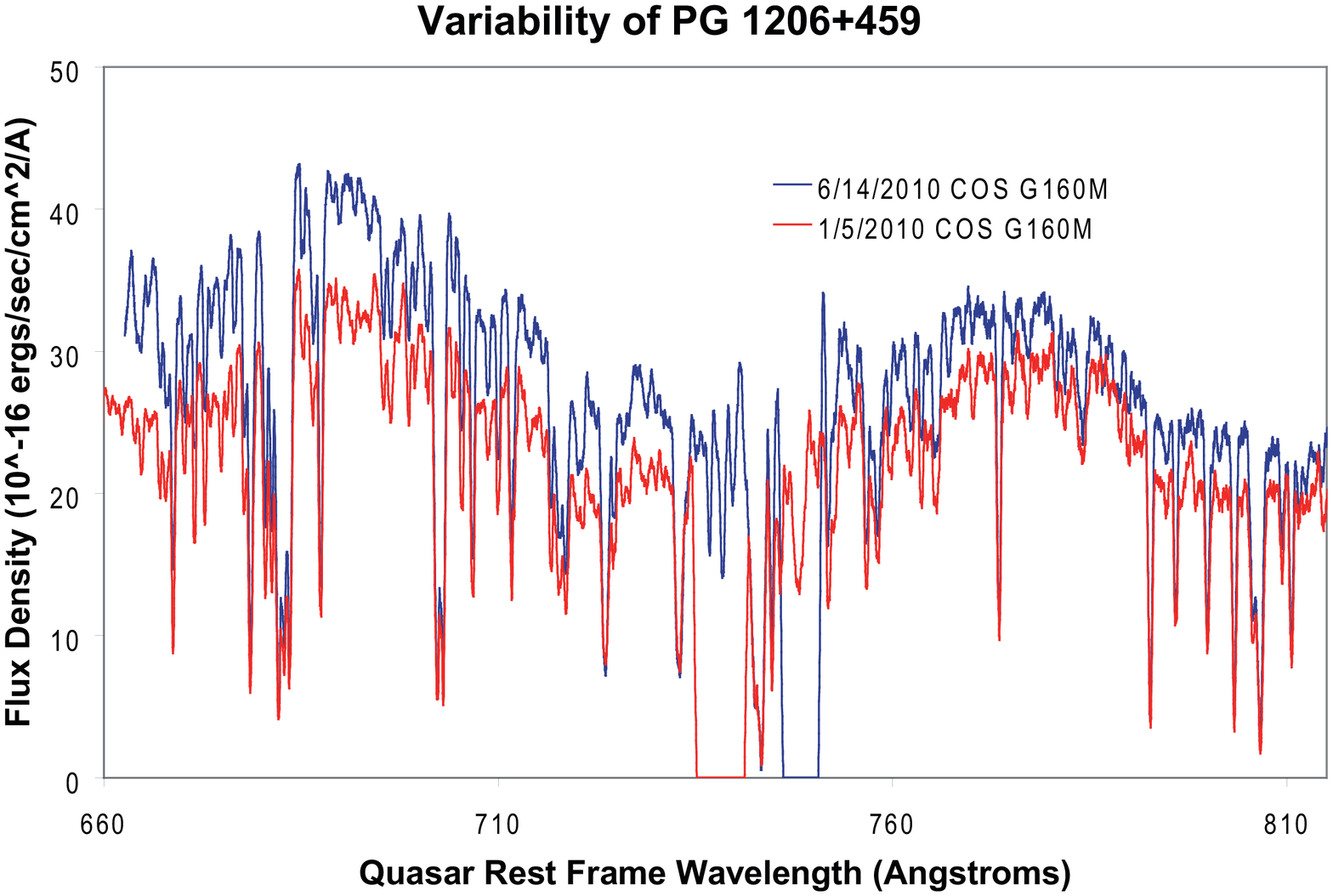}
\includegraphics[width=125 mm, angle= 0]{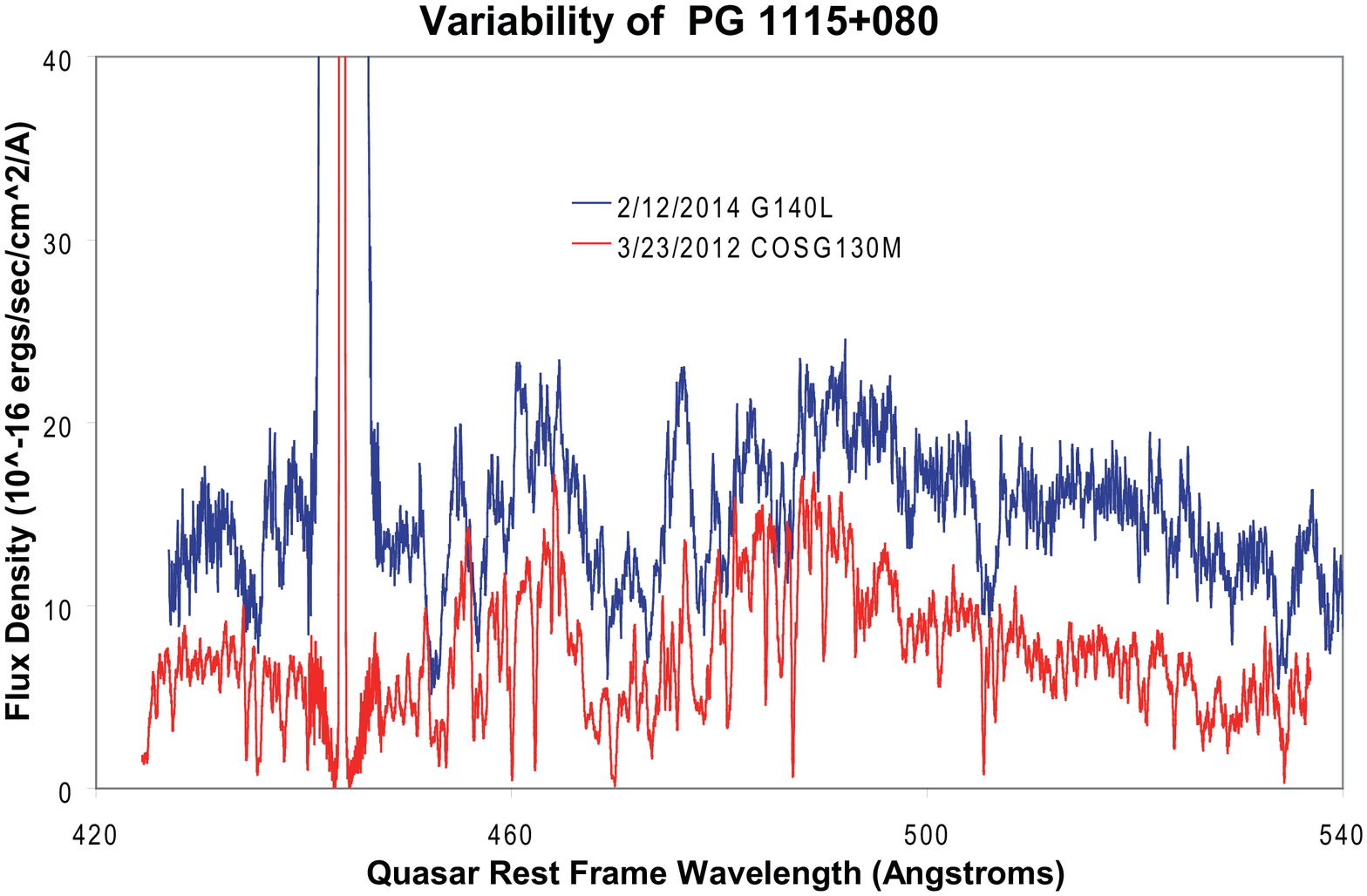}
\caption{Examples of EUV variability detected with HST observations.
The top frame shows modest variability in the quasar PG 1206+459.
The bottom frame is the most variable object in our sample, PG
1115+080. The flux density is the observed value of $F_{\lambda}$.
The wavelengths were changed to the quasar rest frame wavelengths so
as to identify the EUV wavelengths that were sampled. The
$(1+z)^{-3}$ conversion factor of $F_{\lambda}$ to the quasar rest
frame is wavelength independent so it is irrelevant for showing the
degree of variability. Even though there are strong broad emission
lines in PG 1206+459, it is clear that the continuum is varying
significantly between epochs. This example demonstrates that one can
achieve a good estimate of the continuum variability when there are
just a few strong emission lines if one has access to high quality
spectra.}
\end{center}
\end{figure}
\par FOS post-COSTAR aperture positions and alignments were known much more
precisely.  An aperture larger than 0.3" with a ACQ/PEAK during the
target acquisition should yield an accurate flux measurement. If no
ACQ/PEAK was performed, the 0.12" centering uncertainty equates to a
2\% flux uncertainty in a 1.0" circular aperture (FOS Instrument
Science Reports CAL/FOS-140) and this is the minimum aperture that
is chosen for reliable flux measurements. Since the alignments were
well known, post-COSTAR, images (if they exist) can also be used to
determine the centering of the source in the aperture. At the end of
the acquisition sequence some observers took an ACQ image. However,
we were unable to verify accurate centering for any of the
potentially relevant observations with this method when no ACQ/PEAK
step was performed. We were only able to find a few quasar FOS
observations with an adequate aperture size in both the first and
second epoch separated by less than one year in the quasar rest
frame. We also found one second epoch observation with the HRS
spectrograph and a 2.0" aperture (Post-COSTAR) for PG 1008+133.
According to GHRS Instrument Science Report No. 79, the
BRIGHT=RETURN (RTB) feature that was chosen for the spiral search
will automatically return the spacecraft pointing to the dwell
position with the most counts in the aperture. This combined with
the 2.0" aperture should minimize slit losses. Based on this
centering, the uncertainty in the absolute flux should be at most
10\% - 15\% \citep{hea95,sod97}.
\par After the initial experimental stage of testing the COS (Cosmic
Origins Spectrograph), that ended in late September 2009, the flux
calibrations are reliable. Calibration routines are updated based on
the change in sensitivity of the instrument (eg. see Instrument
Science Report COS 2011-02), thus systematic errors are minimized.
The 2.0" slit is suitable for proper alignment as well. Thus, these
observations represent the bulk of our HST sample. The Space
Telescope Imaging Spectrograph (STIS) generally implemented narrow
apertures in order to resolve absorption features and these types of
observations were not intended for accurate flux calibrations.
Slight offsets in the centering of the object would result in loss
of light that would mimic variability. Thus, in spite of numerous
potential observations, STIS was deemed unsuitable for this work for
the potential observations of quasars that we found.

\begin{table*}
 \centering\tiny
\caption{HST Observations of the EUV Continuum of Quasars}
\begin{tabular}{lccccccccccc}
 \hline
 Source     &    z      &    \multicolumn{2}{c}{Date}                       &    T\tablenotemark{a}     &    \multicolumn{5}{c}{Instrumental Setup}                                     &         \\ \cline{3-4} \cline{6-10}
            &           &   Epoch 1 &   Epoch 2 &           &     \multicolumn{2}{c}{Epoch 1   }                    &   &     \multicolumn{2}{c}{Epoch 2   }                 \\\cline{6-7} \cline{9-10}
        &           &           &           &                           &   Spectr./Grating &   Apert.  &   &   Spectr./Grating &   Apert.   \\
    &       &       &       &      [$10^{7}$ s] &       &   ['']    &   &       &   ['']      \\
(1) &   (2) &   (3) &   (4) &   (5) &   (6) &   (7) &   &   (8) &   (9)   \\
\hline
PG 0117+213     &   1.490   &    07/09/1993     &    12/15/1994     &   1.77    &    FOS G160L\tablenotemark{b}         &   4.3 &   &    FOS G270H\tablenotemark{b}         &   0.3  \\
UM 675      &   2.140   &    10/28/1990     &    02/03/1992     &   1.25    &    FOS G160L      &   4.3 &   &    FOS G270H      &   4.3  \\
FIRST   J020930.7-043826    &   1.130   &    12/23/2010     &    1/6/2011   &   0.06    &    COS G130M      &   PSA &   &    COS G230L          &   PSA  \\
SDSSJ   022614.46+001529.7  &   0.620   &    9/22/2010  &    7/29/2011      &   1.65    &    COS G130M  &   PSA &   &    COS G130M  &   PSA  \\
PKSB 0232-042   &   1.440   &    1/9/2010   &    2/16/2010      &   0.13    &    COS G130M  &   PSA &   &    COS G160M      &   PSA  \\
PKSB 0232-042   &   1.440   &    7/31/2015  &    8/8//2015      &   0.03    &    COS E225M          &   PSA &   &    COS E225M  &   PSA  \\
LBQS 0302-0019      &   3.290   &    11/4/1995  &    12/14/1995     &   0.06    &    HRS G140L  &   2.0 &   &    HRS G140L      &   2.0  \\
PG 1008+133     &   1.300   &    11/3/1993  &    6/3/1995   &   2.14    &    FOS G160L\tablenotemark{e}             &   4.3 &   &    HRS G140L\tablenotemark{e}             &   2.0  \\
PG 1115+080A    &   1.740   &    3/23/2012  &    2/12/2014  &   2.13    &    COS G130M              &   PSA &   &    COS G140L              &   PSA  \\
3C 263      &   0.650   &    1/2/2010   &    2/22/2010  &   0.28    &    COS G130M          &   PSA &   &    COS G160M              &   PSA  \\
PG 1148+549     &   0.980   &    12/26/2009     &    12/30/2009     &   0.02    &    COS G130M      &   PSA &   &    COS G130M      &   PSA  \\
PG 1206+459     &   1.160   &    12/29/2009     &    1/5/2010   &   0.04    &    COS G160M          &   PSA &   &    COS G160M          &   PSA  \\
PG 1206+459     &   1.160   &    1/5/2010   &    6/14/2010  &   0.63    &    COS G160M      &   PSA &   &    COS G160M  &   PSA  \\
PG 1329+412     &   1.940   &    08/17/1995     &    06/02/1996     &   0.84    &    FOS G190H\tablenotemark{c}         &   1.0 &   &    FOS G270H\tablenotemark{b}         &   1.0  \\
1334-005    &   2.820   &    02/06/1993     &    06/02/1995     &   1.89    &    FOS G160L\tablenotemark{d}             &   4.3 &   &    FOS G190H\tablenotemark{d}             &   1.0  \\
PG 1338+416     &   1.220   &    5/24/2010  &    5/30/2010      &   0.02    &    COS G130M  &   PSA &   &    COS G160M  &   PSA  \\
PG 1407+265     &   0.940   &    2/12/2010  &    2/21/2010      &   0.04    &    COS G130M          &   PSA &   &    COS G130M      &   PSA  \\
LBQS 1435-0134      &   1.310   &    8/8/2010   &    8/22/2010      &   0.06    &    COS G130M      &   PSA &   &    COS G160M      &   PSA  \\
SDSSJ   161916.54+334238.4      &   0.470   &    8/19/2010  &    6/17/2010      &   1.76    &    COS G130M      &   PSA &   &    COS G140L          &   PSA  \\
PG 1630+377     &   1.480   &    12/13/2009     &    8/01/2010  &   0.78    &    COS G130M      &   PSA &   &    COS G160M          &   PSA  \\
PG 1630+377     &   1.480   &    8/1/2010   &    11/26/2010     &   0.41    &    COS G160M      &   PSA &   &    COS G160M      &   PSA  \\
HS 1700+6416    &   2.740   &    07/26/1994     &    5/29/1996      &   1.52    &    HRS G140L  &   2.0 &   &    HRS G140L          &   2.0  \\
HS 1700+6416    &   2.740   &    10/10/2011     &    4/30/2014      &   2.11    &    COS G140L      &   PSA &   &    COS G140L  &   PSA  \\
HS 1700+6416    &   2.740   &    4/29/2015  &    5/22/2015      &   0.05    &    COS G130M          &   PSA &   &    COS G130M          &   PSA  \\
\hline
\end{tabular}\tiny
\tablenotetext{a}{In quasar rest frame}
\tablenotetext{c}{No peak up, 1.0" aperture Post-COSTAR, both epochs. The 1995 observation showed 2.6\% flux variation between sub-exposures in the sampled range 2230 \AA\ --2280 \AA.}
\tablenotetext{b}{4.3" aperture Pre-COSTAR with peak up andPost-COSTAR 0.3" aperture with peak up.}
\tablenotetext{d}{No peak up first epoch, but a 4.3" aperture. Peakup second epoch, 1.0" aperture Post-COSTAR.}
\tablenotetext{e}{4.3" aperture Pre-COSTAR with peak up and Post-COSTAR 8/1994 2.0" aperture with spiral search for peak flux}
\end{table*}

\begin{table*}
 \centering
\caption{Black Hole Mass and EUV Variability of HST Quasars} \tiny
\begin{tabular}{lccccccccccc}
 \hline
Source  &     $M_\mathrm{bh}$               &    $R_{\rm{Edd}}$     &    T      &            Infall Radius\tablenotemark{a}    &    Variability\tablenotemark{b}       \\
    &       &                       &       &               &       \\
    &     [$10^{9}M_{\odot}$]   &                       &    [M]    &           [M]     &       \\
(1) &   (2) &   (3) &   (4) &           (5) &   (6) \\
\hline

PG 0117+213     &    9.77\tablenotemark{c}  &   0.47    &   388 &           3   &   0.20     \\
UM 675      &    6.95\tablenotemark{d}  &   0.58    &   390 &           3.35    &   0.25     \\
FIRST   J020930.7-043826    &    1.41\tablenotemark{e}  &   0.52    &   92  &           2.3 &   0.05     \\
SDSSJ   022614.46+001529.7  &    0.30\tablenotemark{f}  &   0.68    &   11670   &           14.1    &   0.10     \\
PKSB 0232-042   &    1.48\tablenotemark{g}  &   1.72    &   192 &           6.5 &   0.05     \\
PKSB 0232-042   &    1.48\tablenotemark{g}  &   1.72    &   46  &           3.35    &   0.00     \\
LBQS 0302-0019      &    2.98\tablenotemark{h}  &   2.02    &   58  &           3.85    &   0.05     \\
PG 1008+133     &    2.40\tablenotemark{f}  &   1.02    &   1913    &           9.9 &   0.50     \\
PG 1115+080A    &    4.35\tablenotemark{i}  &   0.64    &   1095    &           5.5 &   1.15     \\
3C 263      &    1.00\tablenotemark{j}  &   0.83    &   591 &           5.75    &   0.18     \\
PG 1148+549     &    1.86\tablenotemark{f}  &   0.61    &   24  &           2.35    &   0.00     \\
PG 1206+459     &    4.79\tablenotemark{f}  &   0.55    &   16  &           2.15    &   0.00     \\
PG 1206+459     &    4.79\tablenotemark{f}  &   0.55    &   282 &           2.85    &   0.33     \\
PG 1329+412     &    2.19\tablenotemark{f}  &   1.36    &   820 &           8.6 &   0.10     \\
1334-005    &    3.06\tablenotemark{k}    &    0.89     &    1320     &            7.1     &   0.15     \\
PG 1338+416     &    1.66\tablenotemark{f}  &   0.72    &   30  &           2.2 &   0.00     \\
PG 1407+265     &    1.41\tablenotemark{e}  &   1.39    &   61  &           4.65    &   0.00     \\
LBQS 1435-0134      &    6.03\tablenotemark{f}  &   0.61    &   20  &           2.2 &   0.00     \\
SDSSJ   161916.54+334238.4      &    0.83\tablenotemark{f}  &   0.22    &   4540    &           4.45    &   0.20     \\
PG 1630+377     &    3.63\tablenotemark{f}  &   1.15    &   463 &           5.8 &   0.00     \\
PG 1630+377     &    3.63\tablenotemark{f}  &   1.15    &   241 &           4.6 &   0.00     \\
HS 1700+6416    &    7.94\tablenotemark{k}    &    1.26     &    411     &            5.8     &   0.05     \\
HS 1700+6416    &    7.94\tablenotemark{k}    &    1.26     &    570     &            6.7     &   0.00     \\
HS 1700+6416    &    7.94\tablenotemark{k}    &    1.26     &    17    &            2.4     &   0.00     \\
\hline
\end{tabular}\tiny
 \tablenotetext{a}{The infall time is equated to a radius based on the slim disk models of Sadowski (2011) that are discussed in Section 5. Assumes a black hole spin of $a/M=0.9$}
\tablenotetext{b}{As computed per Equation(4).}
\tablenotetext{c}{Mass estimated from H$\beta$ profile in
\citet{she03}, using the formula of \citet{asf11}.}
\tablenotetext{d}{Mass estimated from H$\beta$ profile in
\citet{die09}, using the formula of \citet{asf11}. $L_{\rm{bol}}$\
was computed using Equation (3) and the spectrum in \citet{ste91}.}
\tablenotetext{e}{Mass estimated from MgII in \citet{fin14}.}
\tablenotetext{f}{Mass estimated from MgII profile in SDSS spectrum,
using the formula of \citet{tra12}.} \tablenotetext{g}{Mass
estimated from MgII profile in 6df survey spectrum, using the
formula of \citet{tra12}.} \tablenotetext{h}{Mass estimated from
H$\beta$ profile in \citet{syp14}, using the formula of
\citet{asf11}. $L_{\rm{bol}}$\ was computed using Equation (3) and
the SDSS spectrum. see Table 2.} \tablenotetext{i}{See the detailed
discussion in the text} \tablenotetext{j}{Mass estimated from
H$\beta$ profile in \citet{mar03}, using the formula of
\citet{asf11}.} \tablenotetext{k}{Mass estimated from the SDSS CIV
profile using the formula of \citet{par13}.}
\end{table*}

\begin{table*}
 \centering\tabcolsep=3pt
\caption{EUV Variability of SDSS Quasars} \tiny
\begin{tabular}{ccccccccccccc}
 \hline
 Source         &   z   &   \multicolumn{2}{c}{Date}            &   T \tablenotemark{b}    &    Central   $\lambda$            &   \multicolumn{2}{c}{Flux 25 \AA}         &     $M_\mathrm{bh}$    &    Variability  & $R_{\rm{Edd}}$ &  Infall   \\  \cline{3-4}\cline{7-8}
    SDSSJ       &       &      Epoch 1  &     Epoch 2   &       &       &      Epoch 1  &     Epoch 2   &       &                       &     &  Radius\\
                &       &   [MJD]   &     [MJD]     &   [$10^7$\ s] &   [\AA]   &   [10$^{-17}$\ erg s$^{-1}$\ cm$^{-2}$]   &   [10$^{-17}$\ erg s$^{-1}$\ cm$^{-2}$]   &   [$10^{9}M_{\odot}$]   &   & &    [M]     \\
(1) &   (2) &   (3) &   (4) &   (5) &   (6) &   (7) &   (8) &   (9) &   (10)   & (11) &  (12)    \\
\hline  &       &       &       &       &       &       &       &       &     & &         \\
003623.48+242115.3      &   3.586   &   56239   &   56310   &   0.13    &   872.3   &   6140    &   5822    &   1.39 &   0.05   & 0.64 & 2.80 \\
004240.65+141529.6      &   3.724   &   51817   &   53242   &   2.61    &   915.0   &   8143    &   10440.0 &   1.90 &   0.28 & 0.74 & 8.60   \\
011521.20+152453.4      &   3.425   &   54439   &   55830   &   2.72    &   898.3   &   21782   &   11089   &   2.72 &   0.96 & 0.97 &  9.42   \\
021429.29-051744.8  &   3.977   &   55924   &   56662   &   1.28    &   884.2   &   8295    &   7564    &   1.39 &   0.10  & 1.25 & 10.90  \\
030449.84-000813.4 \tablenotemark{a}     &   3.288   &   51817   &   51873   &   0.11    &   905.0   &   61929   &   54288   &   2.98 &   0.14   & 2.02 & 4.02  \\
030449.84-000813.4\tablenotemark{a}     &   3.288   &   51914   &   53378   &   2.95    &   905.0   &   58113.6 &   40073   &   2.98 &   0.45   & 2.02 & 4.26   \\
030449.84-000813.4\tablenotemark{a}     &   3.288   &   51873   &   51914   &   0.08    &   905.0   &   54288   &   58113.6 &   2.98 &   0.07   & 2.02  & 15.10    \\
074212.88+451221.4  &   3.545   &   55483   &   56325   &   1.60    &   896.6   &   10006   &   9585    &   2.39 &   0.04      & 0.76 &  7.05\\
075508.96+413142.0  &   3.442   &   55210   &   56220   &   1.96    &   916.2   &   7069    &   12368   &   2.20 &   0.43      & 0.76 & 7.21 \\
081752.09+105329.6      &   3.327   &   54149   &   55574   &   2.85    &   917.5   &   31391   &   32716   &   2.83 &   0.04  & 1.00 &  10.1  \\
100800.07+222755.1      &   3.454   &   56034   &   56273   &   0.46    &   858.8   &   21062   &   22508   &   1.95 &   0.07   & 0.82 &  4.95 \\
103127.72-012917.1      &   3.371   &   55241   &   55273   &   0.06    &   892.2   &   14899   &   18055   &   2.03 &   0.21   & 0.80 &  2.31   \\
103326.57-020250.7      &   3.375   &   55241   &   55273   &   0.06    &   914.3   &   7722    &   6166    &   1.82 &   0.25    & 0.75 &  2.32  \\
104209.81+060433.4      &   3.535   &   55689   &   55928   &   0.46    &   904.1   &   7048    &   7504    &   1.38  &   0.06   & 0.57 & 3.77 \\
112640.91+380323.6      &   3.805   &   55617   &   55673   &   0.10    &   905.3   &   8566    &   9172    &   1.99 &   0.07   & 0.61 & 2.38  \\
120456.05+234847.3      &   3.579   &   54484   &   56069   &   2.99    &   4.57    &   12538   &   12532   &   2.19 &   0.00   & 1.02 & 9.90  \\
122334.18+052244.7  &   3.342   &   54509   &   55679   &   2.33    &   900.6   &   16318   &   10724   &   2.00 &   0.52    & 0.77 & 8.67 \\
123854.53+193648.6      &   3.372   &   56045   &   56088   &   0.08    &   908.1   &   18451   &   20184.6 &   1.59 &   0.09  & 1.13 & 3.53   \\
123854.53+193648.6      &   3.372   &   54481   &   56045   &   3.09    &   908.1   &   14979   &   18451   &   1.59 &   0.23  & 1.13 &  11.30 \\
124656.95+184926.8      &   3.982   &   54481   &   56090   &   2.79    &   883.2   &   17998   &   17368   &   3.05 &   0.04  & 1.52 & 5.81 \\
130213.54+084208.6      &   3.319   &   54504   &   55978   &   2.95    &   905.2   &   8509    &   6012    &   1.50 &   0.42  & 0.78 &  10.20  \\
131108.71+080959.0      &   3.646   &   54507   &   55958   &   2.70    &   904.0   &   11799   &   14569   &   2.19 &   0.23  & 0.64 & 8.12 \\
131426.03+070301.0      &   3.764   &   54507   &   55958   &   2.63    &   902.7   &   9455    &   14429   &   1.51 &   0.53  & 1.20 &  12.80  \\
131453.02+080456.7      &   3.720   &   54507   &   55958   &   2.66    &   911.0   &   13122   &   11216   &   2.95 &   0.17  & 0.89 & 9.10  \\
133040.39+355525.9      &   3.785   &   54115   &   55603   &   2.69    &   919.6   &   23695   &   24119   &   3.34 &   0.02   & 0.94 & 9.15  \\
135112.50+180922.9      &   3.684   &   54508   &   56038   &   2.82    &   918.1   &   7607    &   11889   &   1.36 &   0.56   & 0.70 & 9.00 \\
140012.85+313452.7  &   3.307   &   54115   &   55276   &   2.33    &   917.1   &   7027    &   8599    &   1.02  &   0.22     & 0.74 & 10.50\\
140509.25+220844.1      &   3.340   &   54537   &   56065   &   3.04    &   910.1   &   16085   &   11683   &   1.87 &   0.38  & 0.67 & 9.25   \\
140644.47+204304.9      &   3.930   &   54527   &   56065   &   2.70    &   892.5   &   14328   &   17965   &   1.67 &   0.25   & 1.85 & 17.20 \\
141458.94+174755.9      &   3.780   &   54523   &   56015   &   2.70    &   873.4   &   7765    &   9225    &   2.17 &   0.19   & 0.66 & 6.80  \\
142437.72+314527.5      &   3.392   &   54252   &   55649   &   2.75    &   905.1   &   6588    &   8107    &   1.22 &   0.23   & 0.56 & 10.40  \\
142519.51+165843.2      &   3.637   &   54506   &   56035   &   2.85    &   900.5   &   7320    &   8222    &   1.25 &   0.12   & 0.86 & 12.20 \\
144752.68+145425.0      &   3.341   &   54507   &   56034   &   3.04    &   898.5   &   17841   &   14043   &   2.16 &   0.27  & 0.75 &  8.60  \\
145344.51+045645.8  &   3.320   &   54560   &   55706   &   2.29    &   914.5   &   7143    &   5456    &   1.22  &   0.31  & 0.61 & 8.10  \\
145646.48+160939.3      &   3.657   &   54535   &   56030   &   2.77    &   891.2   &   7836    &   7995    &   2.49 &   0.02 & 0.57 &  6.33   \\
150810.71+152155.4      &   3.647   &   54535   &   56017   &   2.76    &   903.8   &   15702   &   15472   &   2.11 &   0.01 & 0.91 & 10.30   \\
151754.32+281432.1  &   3.316   &   54173   &   55302   &   2.26    &   904.9   &   11406   &   16163   &   1.28 &   0.42 & 1.14 &   9.52  \\
151756.18+051103.5  &   3.541   &   54562   &   55708   &   2.18    &   896.5   &   13639   &   12472   &   2.95 &   0.09  & 0.90  & 7.87  \\
151929.58+023727.9  &   3.591   &   54560   &   55635   &   2.02    &   914.8   &   11312   &   15145   &   1.39 &   0.34  & 1.04 & 11.90   \\
152856.27+150452.4      &   3.397   &   54243   &   55738   &   2.94    &   909.8   &   13295   &   12948   &   2.46 &   0.03  & 0.90 &  9.52 \\
153303.54+064032.9      &   3.457   &   54208   &   55735   &   2.96    &   913.2   &   15543   &   16689   &   2.20 &   0.07  & 0.81 & 9.56  \\
154806.82+260003.2      &   3.462   &   53846   &   55332   &   2.88    &   919.0   &   8577    &   10896   &   2.07  &   0.27 & 0.61 &  8.60   \\
155644.12+240110.3      &   3.394   &   53786   &   55321   &   3.02    &   919.5   &   10120   &   10223   &   2.61 &   0.01 & 0.62 & 6.52    \\
155812.68+102906.1  &   3.800   &   54572   &   55721   &   2.07    &   916.7   &   6703    &   8991    &   1.47 &   0.34  & 0.66 &  8.85  \\
160521.54+134205.9  &   3.303   &   54568   &   55660   &   2.19    &   906.3   &   15505   &   14055   &   1.95 &   0.10  & 0.86 & 7.92   \\
161030.05+205651.9      &   3.708   &   53793   &   55332   &   2.82    &   920.0   &   9346   &   12943   &   1.96 &   0.39  & 0.85 & 10.30   \\
161624.06+122650.9  &   3.394   &   54571   &   55681   &   2.18    &   910.3   &   7444    &   7554    &   1.72 &   0.01  & 0.45 & 5.85   \\
161625.88+173314.6  &   3.359   &   54585   &   55365   &   1.55    &   917.6   &   13841   &   12504   &   1.86 &   0.11  & 0.82 & 4.52  \\
161625.88+173314.6      &   3.359   &   55365   &   55663   &   0.59    &   917.6   &   12504   &   13989.9 &   1.86 &   0.12 & 0.82 &  6.90   \\
162638.05+215315.6      &   3.351   &   55358   &   55450   &   0.18    &   896.4   &   6304    &   6785    &   1.40  &   0.08  & 0.43 &  2.74  \\
171248.34+371459.7      &   3.350   &   55743   &   56069   &   0.65    &   913.4   &   6800    &   7754    &   1.99  &   0.17 & 0.46 & 4.55    \\
231627.55+225650.4      &   3.494   &   56209   &   56273   &   0.12    &   912.4   &   7939    &   8137    &   2.24 &   0.03  & 0.54 & 2.41   \\
231808.24-003204.5      &   3.330   &   55444   &   55478   &   0.07    &   912.2   &   6015    &   5447    &   1.88 &   0.10  & 0.52 &  2.28  \\
234522.19+151217.3  &   3.599   &   56236   &   56598   &   0.68    &   885.0   &   8125    &   7144    &   1.56 &   0.14    & 0.60 & 5.10 \\
\hline
\end{tabular}
\tablenotetext{a}{LBQS 0302-0019. the mass estimate is from Table
2.} \tablenotetext{b}{Time interval in the quasar rest frame}
\end{table*}
\par In Table 1
we list all of the pairs of epochs of quasar observations (separated
by less than one year in the quasar rest frame) that we believe are
likely to have good calibrations. We also arbitrarily pick a minimum
time interval of 2 days in the quasar rest frame in order to clearly
distinguish separate observations (some observations span many
orbits). We also note that the sample in this section and the SDSS
sample in the next section might have three observations for an
individual quasar. We always pair the observational epochs
consecutively in time and exclude pairing up the first and third
epochs. This procedure ensures the stochastic independence of the
data for subsequent statistical analysis. The first and third
pairing is not independent of the other two consecutive pairings.
Also, a first and third pairing of epochs inordinately weights the
intrinsic variability properties of a single quasar within our
sample. In practice, adding the first and third pairings to our
sample does not affect the analysis and scatter plots to follow
primarily because the one year cutoff prevents long term duty cycle
behavior from canceling out the variability. Excluding the first and
third pairings is formally the best procedure from a statistical
point of view.

\par The data is split between Tables 1 and 2. Table 1 is
organized as follows. Columns (1) and (2) identify the quasar and
the redshift. Columns (3) and (4) are the dates of the two
observations. The next column is this time difference, $T$, in the
rest frame of the quasar in units of $10^{7}$ s. Column (6)
describes the spectrograph and grating used in the first epoch. The
aperture used in the observation is noted in the next column.
Columns (8) and (9) contain the corresponding information for the
second epoch. Table 2 describes the variability time scale in terms
of the estimated central black hole mass. Column (1) is the same as
Table 1. The next column relates this time interval to the size of
the central black hole. We used virial estimates based on the full
width at half maximum (FWHM) in order estimate the mass of the
central black hole, $M_{bh}$ in column (2). The preferred lines are
the low ionization lines MgII or H$\beta$. These lines were
available for all but two high redshift quasars. For these objects
we used a virial mass estimate based on the CIV broad emission line.
Each individual measurement has about 0.3 dex uncertainty (per the
references in Table 2). Thus, each estimate individually is not
accurate enough for our purposes. It is generally believed that by
achieving a sufficient sample size (as in this section and the next
in combination) that the large scatter will be imprinted on a
detectable backdrop of underlying physical trends associated with
black hole mass. By considering the light travel time across the
black hole mass, $M \equiv GM_{bh}/c^{2}$, in geometrized units, we
were able to convert the time interval in column (5) of Table 1 into
a time in geometrized units of $M$ in column (4) of Table 2. In
column (3) of Table 2, we compute the Eddington ratio,
$R_{\rm{Edd}}$,
\begin{equation}
R_{\rm{Edd}} \equiv L_{\mathrm{bol}}/L_{\rm{Edd}}\;,
\end{equation}
where $L_{\rm{Edd}}$ is the Eddington luminosity associated with the
central black hole mass and $L_{\mathrm{bol}}$ is the bolometric
luminosity of the accretion flow. From \citet{pun14}, the luminosity
near the peak of the spectral energy distribution at $\lambda_{e} =
1100$\AA\ (quasar rest frame wavelength), provides a robust
estimator of the bolometric luminosity associated with the thermal
accretion flow, $L_{\mathrm{bol}}$,
\begin{equation}
L_{\mathrm{bol}} \approx 3.8 F_{\lambda_{e}}(\lambda_{e} = 1100
\AA)\;,
\end{equation}
where $F_{\lambda_{e}}(\lambda_{e} = 1100 \AA)$ is the flux density
in the quasar rest frame evaluated at $1100 \AA$.  Note that his
estimator does not include reprocessed radiation in the infrared
from distant molecular clouds. This would be double counting the
thermal accretion emission that is reprocessed at mid-latitudes
\citep{dav11}. If the molecular clouds were not present, this
radiation would be directed away from the line of sight to Earth.
However, it is reradiated back into the line of sight towards Earth
and this combines with the radiation that has a direct line of sight
to Earth from the thermal accretion flow. As such, it would skew our
estimate of the bolometric luminosity of the thermal accretion flow
and needs to be subtracted from the broadband spectral energy
distribution. Because some of these quasars are at high redshift,
there might be Lyman $\alpha$ absorption at $\lambda_{e} =
1100$\AA\, so we compute a similar bolometric correction at
$\lambda_{e} = 1350$\AA\ that follows from the composite SED in
\citet{tel02}
\begin{equation}
L_{\mathrm{bol}} \approx 4.0 F_{\lambda_{e}}(\lambda_{e} = 1350
\AA)\;.
\end{equation}
If possible, we use the average of the two numbers. However, if the
value derived from the spectrum near $\lambda_{e} = 1100$\AA\ is
considerably smaller, we simply used the latter estimate. This
estimate of $R_{\rm{Edd}}$ is used in Section 5 in order to model
the variability in terms of slim disk models. In particular, Column
(5) gives the radial coordinate that is consistent with the infall
of gas to the black hole (in the appropriate slim disk model as
discussed in detail in Section 5 and Figure 8) in a time given by
column (4). The last column is the relative EUV continuum
variability between epochs 1 and 2,
\begin{equation}
V\equiv \rm{Variability}=\frac{\rm{Max}(F1,\, F2)}{\rm{Min}(F1,\,
F2)} -1\;,
\end{equation}
where $F1$ and $F2$ are the EUV fluxes in the first and second
epochs, respectively. For the sources observed with HST, the
variability is computed from the flux density of the EUV continuum
estimated from the entire available spectral overlap between the two
epochs. We do restrict this to below $920\AA$ in order to avoid the
confusion imposed by the very strong broad lines (as discussed in
the next section on SDSS spectra where it is more of an issue). The
process for the HST spectra is to estimate the continuum level in
each HST spectrum individually. These two continuum levels are then
compared in the wavelength region of spectral overlap (below
$920\AA$). We also do not consider noisy edges of the spectra. The
wavelength range used for the the computation of F1 and F2 varies
from epoch to epoch, it depends on the redshift, the gratings that
are used and the noise level at the edge of the spectra. In
practice, below $920\AA$, the variability of the continuum can be
extricated from the emission line variability in the HST spectra
(see for example Figure 1). In the next section, we consider high
redshift SDSS spectra for which there is strong Lyman $\alpha$
absorption and there is no way to fit the EUV continuum from
spectral data. In this case, we use photometry in a wavelength band
that is relatively free of broad emission lines, $850\AA$ -
$920\AA$. The actual wavelength used is listed in Table 3, since
small shifts in wavelength can greatly affect the integrated flux
due to very deep absorption.
\par Figure 1 shows the COS spectra of two of the most variable quasars in the HST sample, PG 1115+080 and PG 1206+459. PG 1115+080
is the most variable quasar in our combined sample and requires a
detailed discussion. PG~1115+080 is a gravitationally lensed quasar.
There are four images, the weaker components B and C and a bright
component A, that resolves into components A1 and A2. A1 is the
located 0.48" from A2 and constitutes 56\% of the total emission at
$7500 \AA$ \citep{tsv10}. The PSA aperture used with these COS
observations has a 2" resolution and cannot resolve component A.
According to Instrument Science Report COS 2010-14, the target
acquisition ACQ/IMAGE should be able to place the peak brightness
associated with the component A in the center of the slit to within
0.1" with very high reliability and there is unlikely significant
light loss. This implies that there is neither confusion over which
lens component is detected nor whether component A was properly
centered in both epochs. It is concluded that the large variability
in Table 2 is real.
\par The second concern with the lensing is possible magnification
that can occur and its effects on $M_{bh}$ estimates and $L_{bol}$.
To minimize this, we use the broad emission line region to make
these estimates. Being much larger than the accretion disk, we
expect that it is less prone to amplification. From the quasar
composite spectrum in \citet{tel02}, one can use the composite
equivalent widths of the broad emission lines in combination with
Equations (2) and (3) and the composite continuum spectral index in
order to achieve a bolometric correction based on line strengths. A
straightforward calculation provides two approximate bolometric
corrections, one based on the MgII line strength and one based on
the CIV line strength, $L(MgII)$ and $L(CIV)$, respectively
\begin{equation}
L_{\mathrm{bol}} \approx 151L(MgII),\quad
L_{\mathrm{bol}}\approx107L(CIV)\;.
\end{equation}
We do not know the relative fraction of the broad emission line
luminosity that originates in component A1 of the lensed quasar,
PG~1115+080. However, as stated above,  A1 produces 56\% of the
total emission detected at Earth at $7500 \AA$ \citep{tsv10}. We
assume, in the absence of any other information that $\approx 56\%$
of the MgII and CIV line strength is also from A1. Thus, from the
estimator in Equation (5), $L_{\mathrm{bol}} = 3.0 \times 10^{47}
\rm{erg/s}$ and $L_{\mathrm{bol}} = 4.0 \times 10^{47} \rm{erg/s}$
for MgII and CIV, respectively.  Equation (5) yields reasonable
agreement between the two estimators based on the two different line
strengths in the case of PG~1115+080. Taking the average, we
estimate that component A1 has $L_{\mathrm{bol}} \approx 3.5 \times
10^{47} \rm{erg/s}$ and this represents $L_{\mathrm{bol}}$ of the
quasar. Similarly, we estimate $M_{bh}$ using the line strength and
not the continuum with the expectation that were are minimizeing the
affects of magnification from the lens. We use the more reliable low
ionization line MgII and the formulas from \citet{she12} to find
\begin{eqnarray}
&&\log\left(\frac{M_{bh}}{M_{\odot}}\right) = 3.979
+0.698\log\left(\frac{L(MgII)}{10^{44} \,\rm{erg/s}}\right) +
1.382\log\left(\frac{\rm{FWHM}}{\rm{km/s}}\right)=9.758\;,
\nonumber\\
&& \frac{M_{bh}}{M_{\odot}} = 5.36 \times 10^{9}\;.
\end{eqnarray}
Alternatively, the formula of \citet{tra12} yields a different
estimate
\begin{equation}
\frac{M_{bh}}{M_{\odot}} = 6.79\times 10^{6}
\left(\frac{L(MgII)}{10^{22}
\,\rm{erg/s}}\right)^{0.5}\left(\frac{\rm{FWHM}}{1000
\,\rm{km/s}}\right)^{2} =3.35 \times 10^{9}\;.
\end{equation}
Taking the average of Equations (6) and (7), we arrive at a mass
estimate, $M_{bh} =4.35 \times 10^{9}M_{\odot} \pm 1.00 \times
10^{9}M_{\odot}$. The associated Eddington ratio of 0.64 in Table 2
is typical for broad absorption line quasars such as PG 1115+080
\citep{gan07}. Any magnification of the broad line region from
lensing would imply that the intrinsic Eddington ratio is actually
smaller. Thus, we expect that the magnification is at most a factor
of 2 or 3 (yielding an Eddington ratio of 0.35 - 0.4) in order for
PG 1115+080 to be consistent with typical Eddington ratios found in
broad absorption line quasars. We acknowledge that there is
additional uncertainty associated with the $M_{bh}$ and
$R_{\rm{Edd}}$ for PG 1115+080 due to the gravitational lensing.
Fortunately, the results of this paper do not depend on these
estimates for PG 1115+080. The trends persist even without this
quasar. We choose to keep it in our sample since the discussion
above indicates that the large variability is not an artifact of the
gravitational lensing and a centering uncertainty in the apertures.
\par The sample is composed predominantly of radio
quiet quasars. There are four radio loud quasars. Two of them are
extended quasars with large steep spectrum radio lobes, 3C 263 and
PKSB 0232-042. The EUV spectra were analyzed by us previously
\citep{pun15}. The other two are the unresolved flat spectrum radio
sources, FIRST J020930.7-043826 and UM 675. Considering the strong
EUV emission lines present in the HST spectrum, we do not expect
significant blazar contamination of the continuum in FIRST
J020930.7-043826 \citep{fin14}. The spectral energy density of UM
675 is increasing slightly across the near UV with strong broad
lines with a spectral energy density two orders of magnitude higher
than the microwave luminosity and is therefore likely dominated by
the accretion disk in the EUV \citep{ste91}. In all four cases, the
variability in the EUV spectra should be indicative of the thermal
accretion flow.

\section{SDSS Extreme Ultraviolet Variability Data} The SDSS
archive offers a large database for quasar spectra and therefore
great potential for a study of EUV variability. The main obstacle
results from the large redshifts that are required to shift the
quasar EUV continuum into the optical window. Even a moderate
sensitivity study (say a threshold of 20\% change as a detection for
the sake of argument) requires decent signal to noise spectra. High
redshift quasars have diminished fluxes due to the large distances,
but also the intervening Ly$\alpha$ absorbing clouds greatly
attenuates the signal in the EUV. The other issue that exacerbates
the situation is the desire to monitor the EUV continuum and not the
more distant emission line region. Thus, we must separate out the
broad emission lines from the continuum. The long wavelength end of
the EUV continuum, with the least Ly$\alpha$ absorption and highest
fluxes in general, is contaminated by strong, extremely broad
emission lines, OVI, Ly$\beta$ and the blend of Ly$\gamma$, SVI,
CIII and NIII. In order to sample the EUV shortward of these strong
emission lines requires sampling the rest frame spectrum below
$\lambda_{e} =920\, \AA$. In the range of $850\,\AA$ to $920\, \AA$
there is a region with no strong emission lines. Due to the
Ly$\alpha$ absorption troughs, integration over a finite window is
necessary in order to reliably compare fluxes from epoch to epoch.
In order to accommodate a finite window of integration, the shortest
useable wavelength for early SDSS spectra is $\lambda_{o} = 3850\,
\AA$ (where $\lambda_{o}$ is the observed wavelength). In order to
place, the quasar rest frame wavelength $\lambda_{e}=920\, \AA$ at
$\lambda_{o}
> 3850\, \AA$, requires a redshift, $z > 3.3$. The challenge is the
following. As the wavelength decreases below $\lambda_{e}=1100\,
\AA$, the Ly$\alpha$ absorption from intervening gas, more often
than not, increases as the wavelength decreases. This decreases the
signal to noise of the spectrum at the shortest wavelengths. In most
cases, the signal to noise ratio is too low for accurate variability
studies. Thus, we attempt to find the shortest wavelength that has
adequate signal to noise with the constraint that the observed
wavelength corresponds to $\lambda_{e}<920\, \AA$. Alternatively
stated, the quasars must be bright and the Ly$\alpha$ absorption not
particularly deleterious. This eliminates most high redshift
quasars.

\par Thusly motivated, our search criteria was all quasars in the
redshift range $3.3 < z < 4.0$ with two epochs of observation that
are separated by less than one year in the quasar rest frame. We
institute an ``integration window" of $\Delta\lambda_{e} = 25\, \AA
$, which corresponds to $\Delta\lambda_{o} \approx 110\, \AA $. The
window minimizes any affects caused by wavelength calibration and
sky subtraction differences between observations and the many narrow
absorption lines, allowing for a consistent method of determining
the observed flux density. We started with 835 potential pairs of
observations (with a time separation in the quasar rest frame
between 2 days and 1 year) of which the 54 pairs of observation that
had sufficient flux in both epochs to qualify for the sample are
listed in Table 3.
\par A minimum signal to noise ratio needs to be found in order to
eliminate the possibility of random noise creating numerous false
positive variations that mask any structure in the temporal
dependence of variability. A basic test is that for the very
smallest intervals, on the time scales of days in the quasar rest
frame, the variations should be distributed near zero with a
``small" standard deviation. In the $\Delta\lambda_{e} = 25\, \AA $
window, we find that an observed average flux density of $3.0 \times
10^{-17} \rm{erg/s}/\rm{cm}^{2}/\AA$ is a good lower flux density
cutoff. In the blue end of the SDSS spectra, the rms noise is more
of a concern than sky subtraction errors for weak sources. Sky
subtraction errors are evident from the fact that Ly$\alpha$
absorption troughs can extend below zero. We estimate sky
subtraction errors in the blue to be typically $\sim 1.0 \times
10^{-18} \rm{erg/s}/\rm{cm}^{2}/\AA$. Judging from the intensity
below 0 in some of the deep Ly$\alpha$ troughs, the effect can reach
$\sim 1.0 \times 10^{-17} \rm{erg/s}/\rm{cm}^{2}/\AA$ in a few
cases. The worst case rms noise near $\lambda_{0} = 3850\, \AA $ is
$\rm{RMS} < 1.0 \times 10^{-17} \rm{erg/s}/\rm{cm}^{2}/\AA$. Thus,
an observed average flux density of $3.0 \times 10^{-17}
\rm{erg/s}/\rm{cm}^{2}/\AA$ has a signal to noise ratio,
$\rm{SNR}>3$ and more typically one has $\rm{SNR}\approx 5$. The SNR
over the range of $\Delta\lambda_{e} = 25\, \AA $ (equivalent to a
window of $\Delta\lambda_{o} \approx 110 \, \AA $) will be $\rm{SNR}
\sim  110 \,\rm{RMS} / \sqrt{110}\,\times 10^{-17} \sim (110) \, 3.0
\times 10^{-17}/ \sqrt{110}\times 10^{-17} \sim \sqrt{110}
\,\rm{RMS}> 30$. Thus, random noise should exceed $10\%$ of the
average flux density with a probability of less 0.01. Thus, random
noise should not be an issue in our variability analysis, leaving
systematic affects (such as large sky subtraction errors) as the
main source of false positive variability.
\begin{figure}
\begin{center}
\includegraphics[width=80 mm, angle= 0]{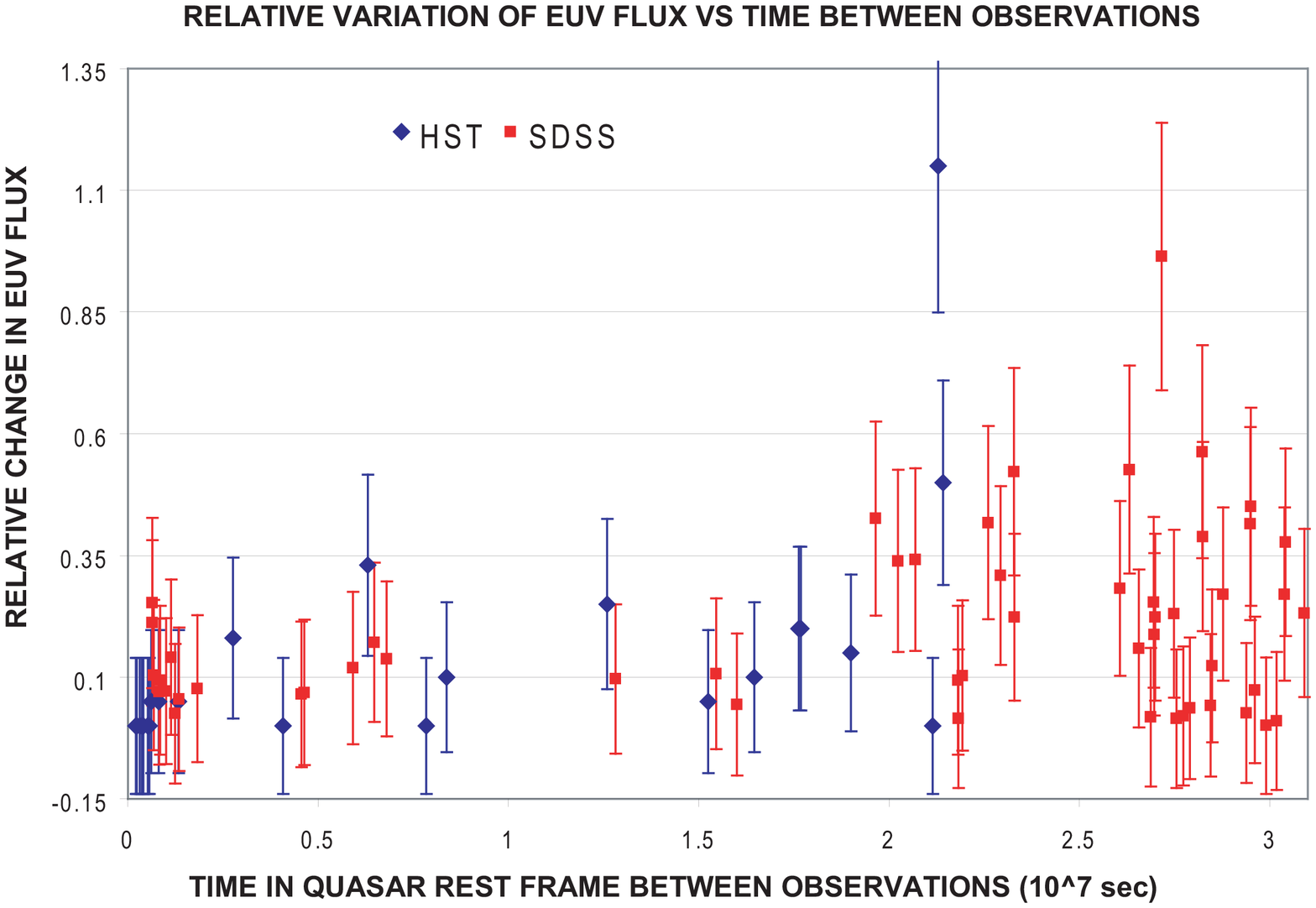}
\includegraphics[width=80 mm, angle= 0]{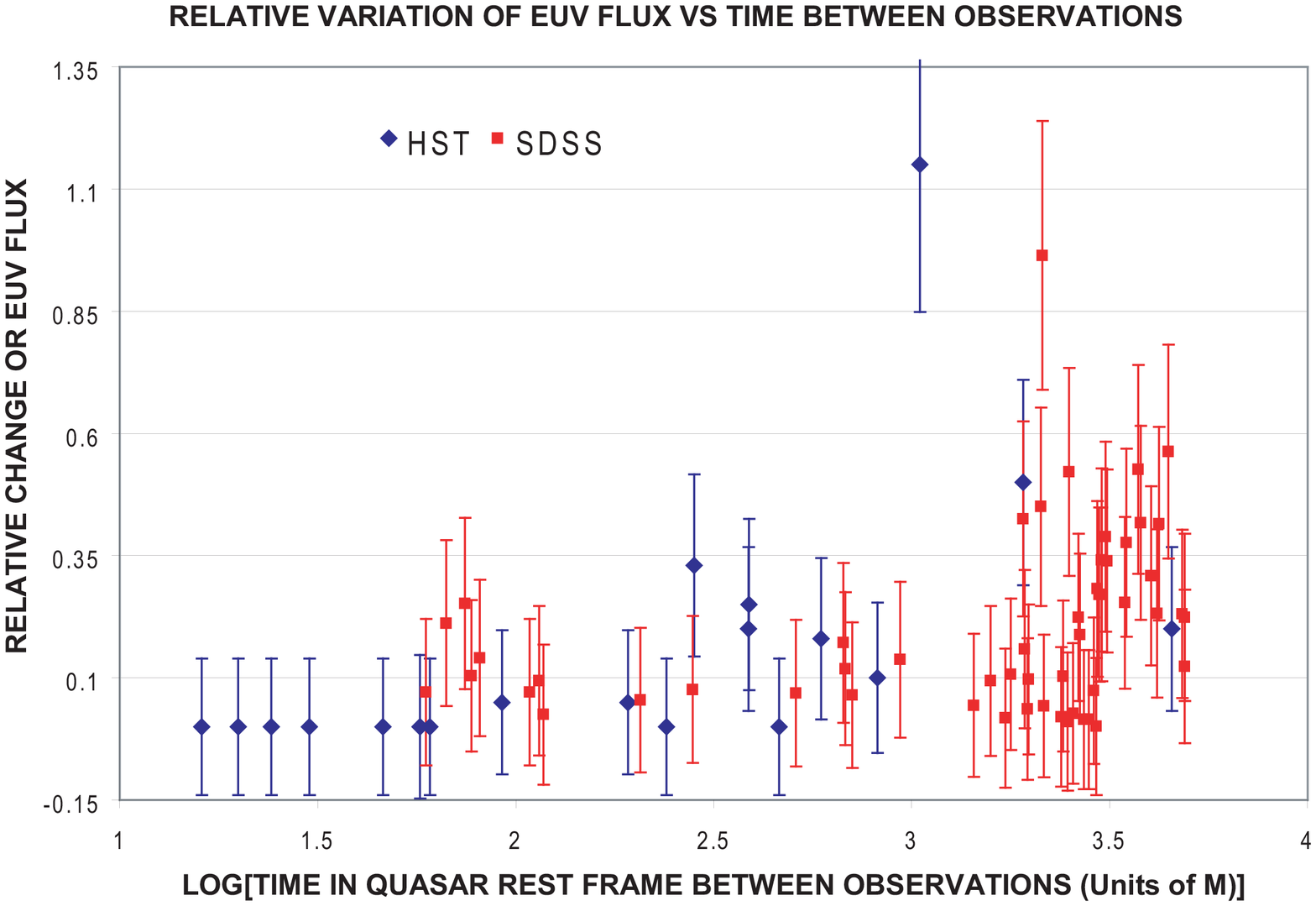}
\caption{The left hand frame of Figure 2 combines the data in Tables
1 - 3. It is a scatter plot of the EUV variability computed with
Equation (4) versus the time between individual epochs of
observation. The variability is clearly larger for time intervals
$>2\times 10^{7}$ s compared to time intervals $<1.5\times 10^{7}$
s. The right hand frame converts this time to geometrized units of
$M$. The variability is clearly larger for time intervals $>1800$M
compared to time intervals $<1000$M. The error bars are derived from
an $\approx 10\%$ uncertainty in each absolute flux measurement as
discussed in the text.}
\end{center}
\end{figure}
\begin{figure}
\begin{center}
\includegraphics[width=80 mm, angle= 0]{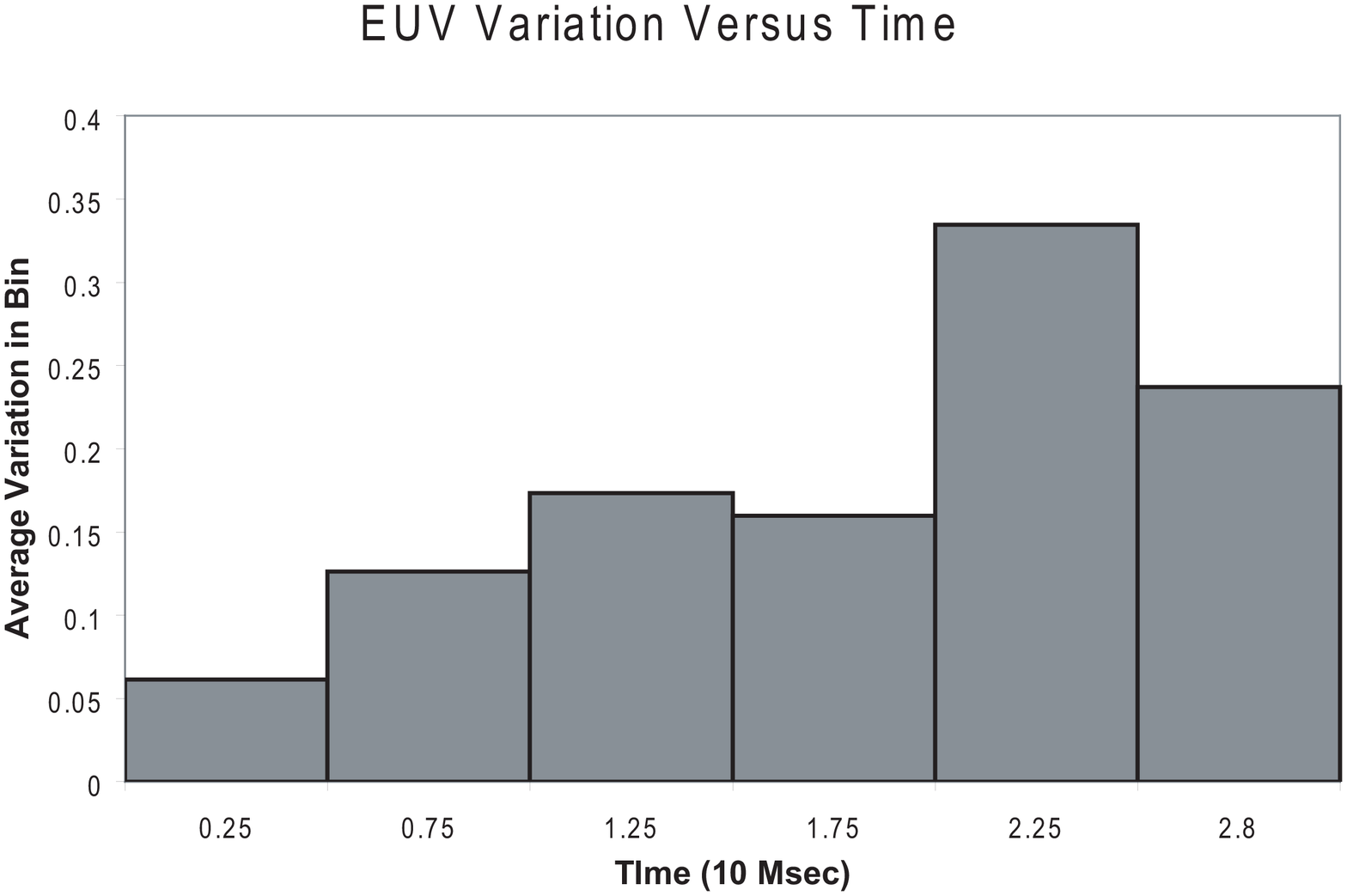}
\includegraphics[width=80 mm, angle= 0]{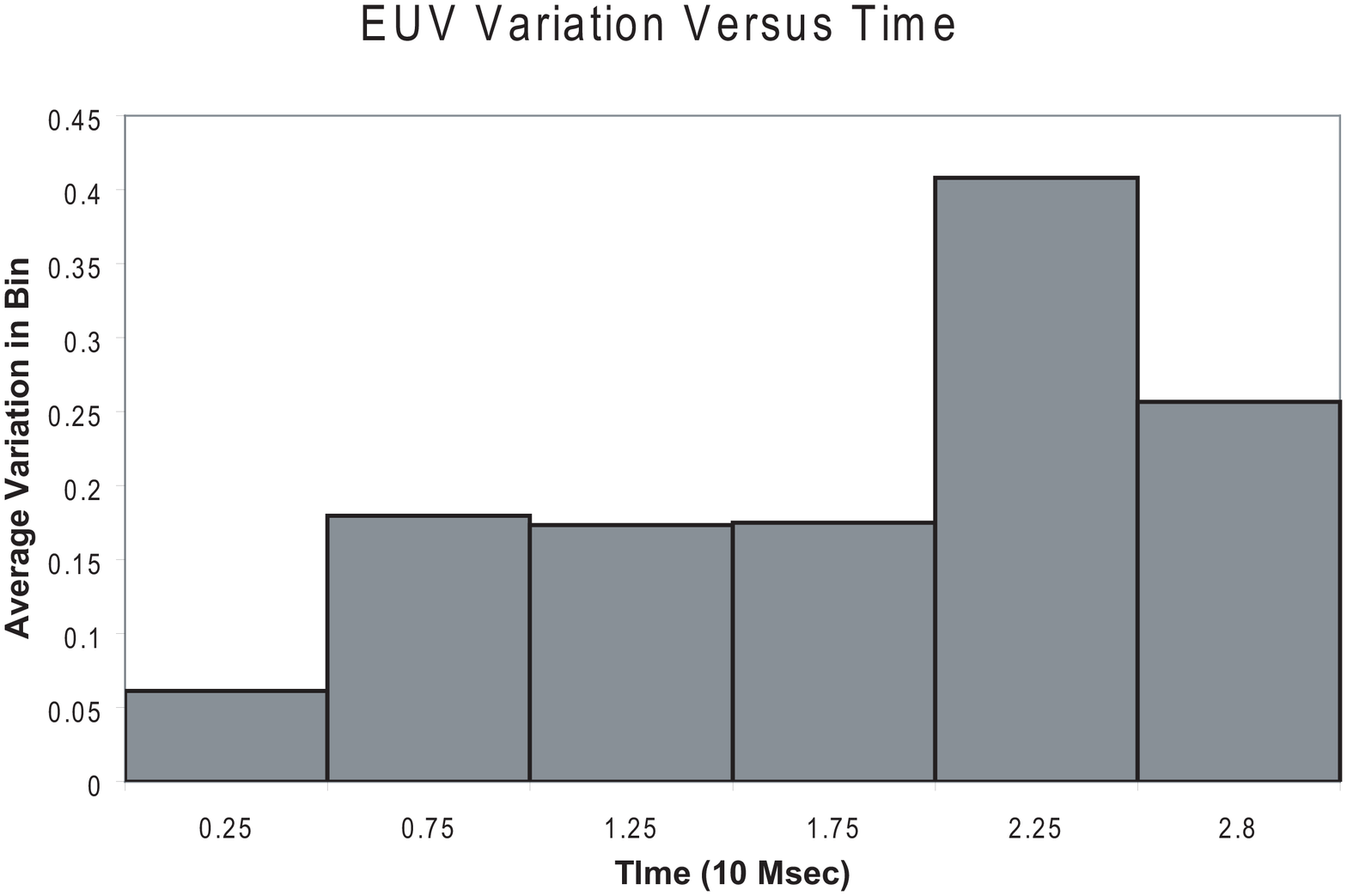}
\caption{The left hand frame is a histogram of the variation of the
data in Figure 2 versus time. The value in each bin is the average
value in the bin. The right hand frame shows the effect of reducing
the average signal to noise required for inclusion in the sample of
Table 3 from $3.0 \times 10^{-17} \rm{erg/s}/\rm{cm}^{2}/\AA$ to
$2.0 \times 10^{-17} \rm{erg/s}/\rm{cm}^{2}/\AA$. Note that the last
bin is slightly wider, ranging from $2.5 \times 10^{7}$s to $3.1
\times 10^{7}$s.}
\end{center}
\end{figure}

\par There are two main issues that affect the
veracity of SDSS for detecting modest variability. The first was
establishing an acceptable SNR for the data. This was addressed
above. The second is the calibration of quasar spectra that were
observed as part of the Baryon Oscillation Spectroscopic Survey
(BOSS) \citep{daw13}. For BOSS quasar data, there is a calibration
issue in the blue part of the spectrum. Specifically, in order to
increase spectral sensitivity for quasars, the hole drilled in the
plate over the fiber is centered on the displacement associated with
4000\AA. However, the calibration of the star has a hole centered
about the displacement associated with 5400\AA. Due to atmospheric
differential refraction (ADR), this inconsistency artificially
raises the flux level in the blue for the quasar and underestimates
the flux density of the standard star at 4000\AA\,. The quasar
affect is small, the observed wavelength is $3850\, \AA <
\lambda_{0} < 4150\, \AA$, so the ADR displacement of the light from
the center of the fiber is minimal. However, the calibration
standard star was observed with the fiber at a displacement centered
at 5400\AA\, and there is significant displacement from the center
of the fiber at $\sim 3900\AA$ and these fluxes are significantly
underestimated. These effects can enhance the detected quasar blue
flux by more than 15\%. Pictures of the plate and a discussion of
this circumstance can be found in Dawson et al. (2013). We have
corrected for the BOSS ADR affect in Table 3 with the following
methodology. We retrieved the $50^{\rm{th}}$ percentile (median)
seeing values and the airmass from the header of the spectra. We
then computed the displacement due to ADR between $4000\,\AA$ and
$5400\,\AA$ for atmospheric air pressure at Apache Point (2800 m),
and corrected for the relative enhancement of the $4000\,\AA$ flux
with respect to $5400\,\AA$ flux assuming a 2D Gaussian shape for
the seeing image of the quasar (seen within the 2" aperture of the
BOSS plate holes), and integrating the light loss following the
technique of \citet{fil82} for a circular aperture.

\par The results of our search for EUV variability can be found in
Table 3. The table is organized as follows. The first two columns
are the source and the redshift. The next two columns are the date
of the paired observations. Column (5) is the time interval in the
rest frame of the quasar. Column (6) is the wavelength at the center
of the 25\AA\, window in the quasar rest frame.  We try to make it
as short of a wavelength as possible with the restriction that the
Ly$\alpha$ absorption does not suppress the average flux below our
limit $3.0 \times 10^{-17} \rm{erg/s}/\rm{cm}^{2}/\AA$. We also try
to avoid placing deep absorption troughs near the window edges as
this can increase the potential for systematic errors in integration
from epoch to epoch. We have not optimized this central wavelength.
Columns (7) and (8) are the integrated fluxes in these windows
evaluated in the quasar rest frame. Note that this can be converted
to the average in the observer's frame by dividing by $25\AA
(1+z)^{3}$. This number can fall below our limit of $3.0 \times
10^{-17} \rm{erg/s}/\rm{cm}^{2}/\AA$ because these are BOSS
corrected fluxes. The limit applies to uncorrected fluxes. This is
appropriate since the BOSS observations were designed to increase
signal to noise at the blue end of the spectra at the expense of an
accurate absolute flux calibration. Column (9) is $M_{\rm{bh}}$
estimated from the CIV broad emission line in the SDSS spectrum
\citep{par13}. Each individual measurement has about 0.3 dex
uncertainty \citep{par13}. Thus, each estimate individually is not
accurate enough for our purposes. However, recall the strategy that
we declared in the previous section concerning the HST mass
estimates. It is generally believed that by achieving a sufficient
sample size (as in this section and the previous in combination)
that the large scatter will be imprinted on a detectable backdrop of
underlying physical trends associated with black hole mass. The next
column is the variability computed with Equation (4). Column (11) is
$R_{\rm{Edd}}$. The last column is the infall radius in the slim
disk model associated with $M_{\rm{bh}}$ and $R_{\rm{Edd}}$ for
which the infall time (computed in Section 5) equals the time
interval between observations in column (5).
\par Our criteria of an observed average flux
density of $>3.0 \times 10^{-17} \rm{erg/s}/\rm{cm}^{2}/\AA$,
eliminates random noise as likely contributor to false positive
identifications of variability at $>10\%$ level. This leaves
systematic affects as the most likely cause of potential false
positive identification of variability. We explore the propensity of
such behavior in the SDSS database by utilizing an empirical
calibration method based on field galaxies \citep{yip09}. Using the
assumption that the background field galaxies should not be
variable, we derive a wavelength dependent correction. The
correction varies across the plate, so it is only valid in a small
region near each quasar. In general two plates cannot be compared
unless they are identical plates at two different dates. Otherwise,
we find that there are not enough overlapping galaxies with spectra
on both plates in small neighborhoods of each quasar. We applied
this method of calibration to our quasars in our EUV centered
windows in order to get an estimate the accuracy of the flux
integrations in out sample. We performed the following steps
\begin{enumerate}
\item A neighborhood of the plate was excised around each quasar, with a right ascension $<2^{\circ}$
from the quasars and a declination $<20$ arc-minutes from the
quasar. The tight constraint on the declination follows from the
fact that declination influences most of the light loss associated
with the ADR since these sources were in general observed at small
hour angle. We found that using the full plate instead of the
contiguous region to the quasar can create errors as large as 15\%.
\item The  individual spectra of galaxies with $\rm{SNR} >3$ were primary and non-primary observations of
the same objects and were downloaded from the SDSS DR12 sky server.
\item By forming a wavelength dependent ratio of the flux densities of the field galaxies in the first epoch to the
those in the second epoch in each quasar neighborhood, we determined
a re-scaling wavelength dependent function that we applied to the
measured quasar fluxes in the second epoch.
\item Unlike \citet{yip09}, it was found that dividing the individual field galaxy spectra
and then summing to retrieve the re-scaling function does not give
stable results. Results are influenced by noise. Stable results are
obtained by adding up the field galaxy spectra of each plate and
then dividing the two plate cumulative spectra in order to obtain
the re-scaling function.
\end{enumerate}
\par We considered the plates with at least 20 field galaxies in
common in the local quasar neighborhood defined in item (1) above.
We then computed the re-scaling function and applied it to the EUV
window for the quasar. The re-scaled flux was $1.02 \pm 0.11$ times
the ``uncorrected" flux, i.e. the fluxes that appear in Table 3. If
we insist on at least 30 galaxies in common between the two plates
the re-scaling factor in the EUV window is $0.97 \pm 0.07$. The
implication is that the SDSS data taken directly from the pipeline
with our SNR minimum in the EUV window is typically accurate to
within 10\%. We verify the magnitude of the systematic uncertainty
by considering the short time durations in Table 3. For short time
durations, we expect virtually no variability in the EUV. For time
intervals $< 2\times 10^{6}$ seconds (less than 3 weeks in the
quasar rest frame), we find that the ratio of the EUV fluxes between
epoch 1 and epoch 2 are $F1/F2 = 1.00 \pm 0.13$ which is consistent
with most of the variation being from the systematic errors found
from our field galaxy re-calibration of a few selected quasars.

\par
A similar method has been used based on background stars instead of
field galaxies \citep{wil05}. We tried to implement this method as
well. However, we were able to find  background stars in common in a
local neighborhood of the quasar only in 3 cases. We preferred to
stay with galaxies that allow the re-scaling of a larger number of
plate pairs.
\section{The Combined Data Sample} In this section, we combine the data
in the HST and SDSS samples to improve the statistical analysis.
This is viable because our selection criteria produced a common bias
that selected very luminous quasars. This is a necessary consequence
for an SDSS sample based on very high redshift quasars that are
sufficiently luminous to shine brightly through the Lyman~$\alpha$
forest. Similarly this is a consequence of the practicality of
allocating HST time for multiple observations of quasars in the EUV.
It would be difficult to obtain telescope time for high resolution
spectroscopy in the EUV for any source that was not very bright in
the EUV. Such observations require one or more full orbits to
accomplish even for the EUV bright quasars in our sample. Thus, we
acquired quasars in both samples with very large $M_{\rm{bh}}$ and
$R_{\rm{Edd}}$ as evidenced by the calculations presented in Tables
2 and 3. In particular, $R_{\rm{Edd}} =0.83\pm 0.32$ and
$R_{\rm{Edd}} =0.95\pm 0.48$ for the SDSS and HST samples,
respectively. No statistical difference can be found in the parent
population with a Kolmogorov-Smirnov test or a Wilcoxon rank sum
test. The median $M_{\rm{bh}}$ are $1.96 \times 10^{9}M_{\odot}$ and
$2.29 \times 10^{9}M_{\odot}$ for the SDSS and HST samples. However,
the variance of the HST sample is much larger, as expressed by the
mean values of $M_{\rm{bh}}= 1.98\pm 0.55 \times 10^{9}M_{\odot}$
and $M_{\rm{bh}}= 3.40\pm 2.61 \times 10^{9}M_{\odot}$. No
statistical difference is found between the two samples in a
Wilcoxon rank sum test since the medians are similar, but the
difference in variances make for a 0.012 probability that the
$M_{\rm{bh}}$ samples are drawn from the same parent sample by
random chance according to a Kolmogorov-Smirnov test. For our
purposes, this is not very important since the $R_{\rm{Edd}}$
distribution is statistically indistinguishable between the two
samples, the larger variance of $M_{\rm{bh}}$ and similar medians
just allows for a wider range of $M_{\rm{bh}}$ to be explored with a
similar $R_{\rm{Edd}}$ as a result of combining the samples. Thus,
it is very appropriate from a statistical analysis point of view to
combine these two samples in the following.

\par The left hand frame of Figure 2 combines the data in Tables 1 -
3. Approximately 30\% of the objects are HST quasars in the redshift
range, $0.5 < z < 1.5$ and 70\% are SDSS quasars with $z\approx
3.5$. It is a scatter plot of the EUV variability computed with
Equation (4) versus the time between individual epochs of
observation. The errors are derived from the analysis and
observation selection criteria of the last two sections. Recall that
each observation has an $\approx 10\%$ uncertainty in the absolute
flux measurement. This results in $\approx 14(1+V)\%$ uncertainty
($V$ is the relative variability that was defined in Equation (4)))
in the variability calculation. The most striking aspect is the
abrupt change in the distribution of the degree of variability when
the time scale exceeds a threshold of $\approx 2 \times
10^{7}\rm{s}$. Significant variations are much more common above
this threshold. More precisely stated, the variability is clearly
larger for time intervals $>2\times 10^{7}$ s compared to time
intervals $<1.5\times 10^{7}$ s. We try to explore the relevance of
this visual appearance in terms of sampling density and measurement
uncertainty in the remainder of this section. This is a rather
startling find because there is a range of $M_{\rm{bh}}$ and
$R_{\rm{Edd}}$ (although rather narrowly distributed as discussed in
the introductory paragraph of this section) evident in Tables 2 and
3. In order to ``normalize" these times, we convert the times into
geometrized units. The right hand frame shows the scatter plot, with
the time converted to units of light travel time across the central
black hole mass, $M$ in geometrized units. The transition to
variability looks abrupt in terms of geometrized units as well.
There appears to be an abrupt change in the distribution of the
degree of variation that occurs at times $> 1800 M$ in the right
hand frame of Figure 2.
\begin{figure}
\begin{center}
\includegraphics[width=80 mm, angle= 0]{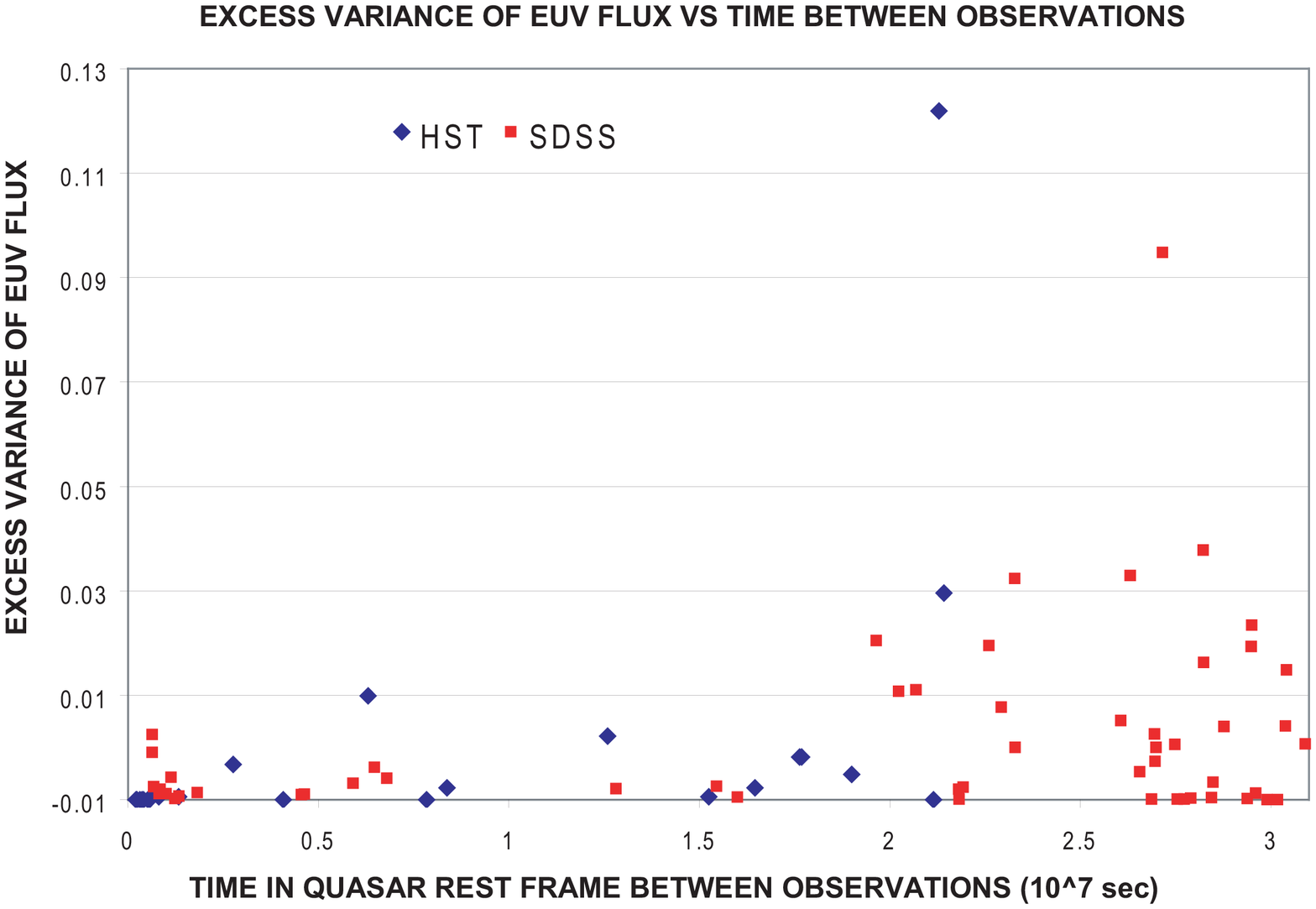}
\includegraphics[width=80 mm, angle= 0]{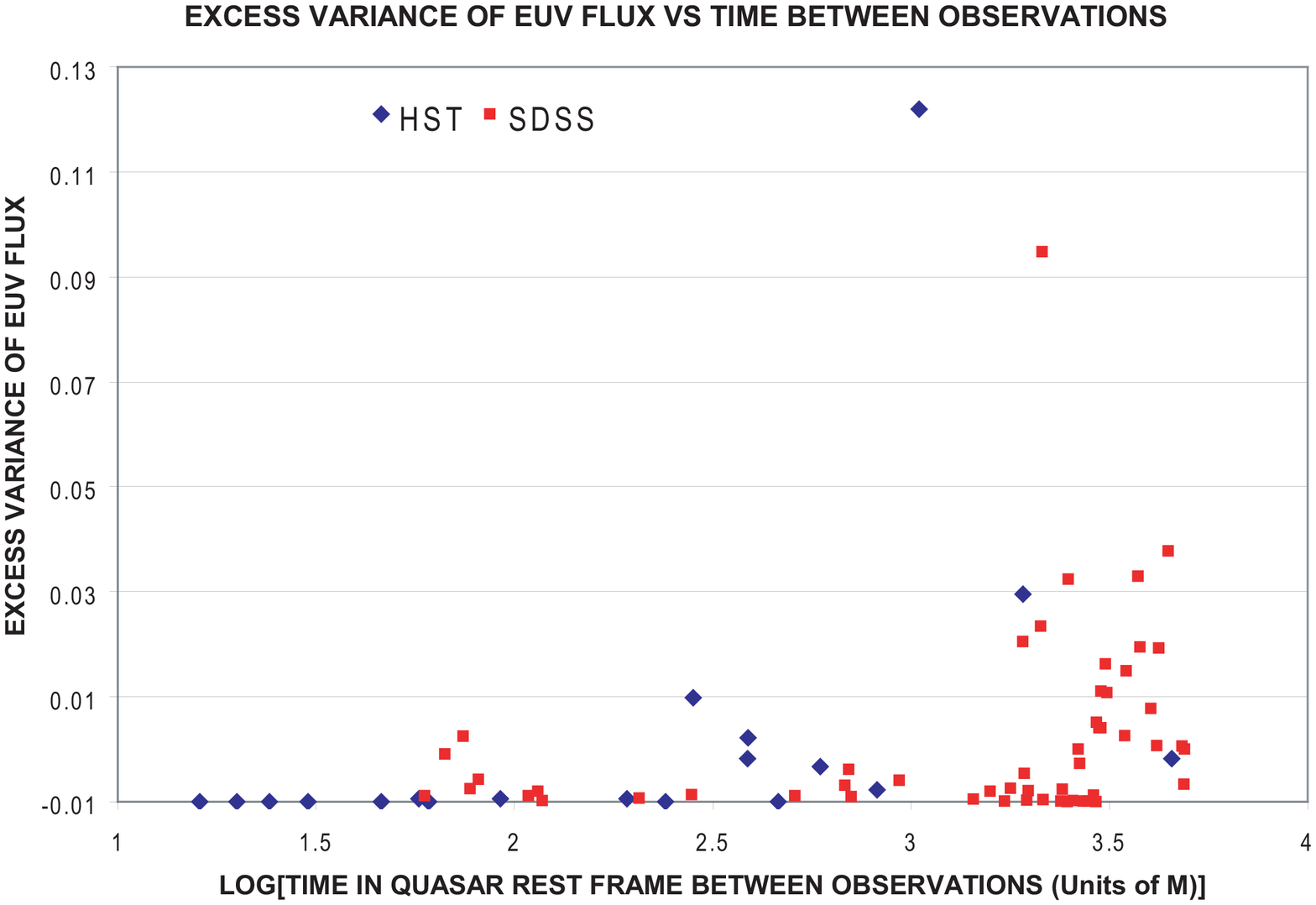}
\caption{The left hand frame is a scatter plot of the excess
variance of the paired quasar observations versus the time between
the observations in the quasar rest frame. The right hand frame is
the same with the time converted into units of $M$, the black hole
mass in geometrized units.}
\end{center}
\end{figure}
\begin{figure}
\begin{center}
\includegraphics[width=80 mm, angle= 0]{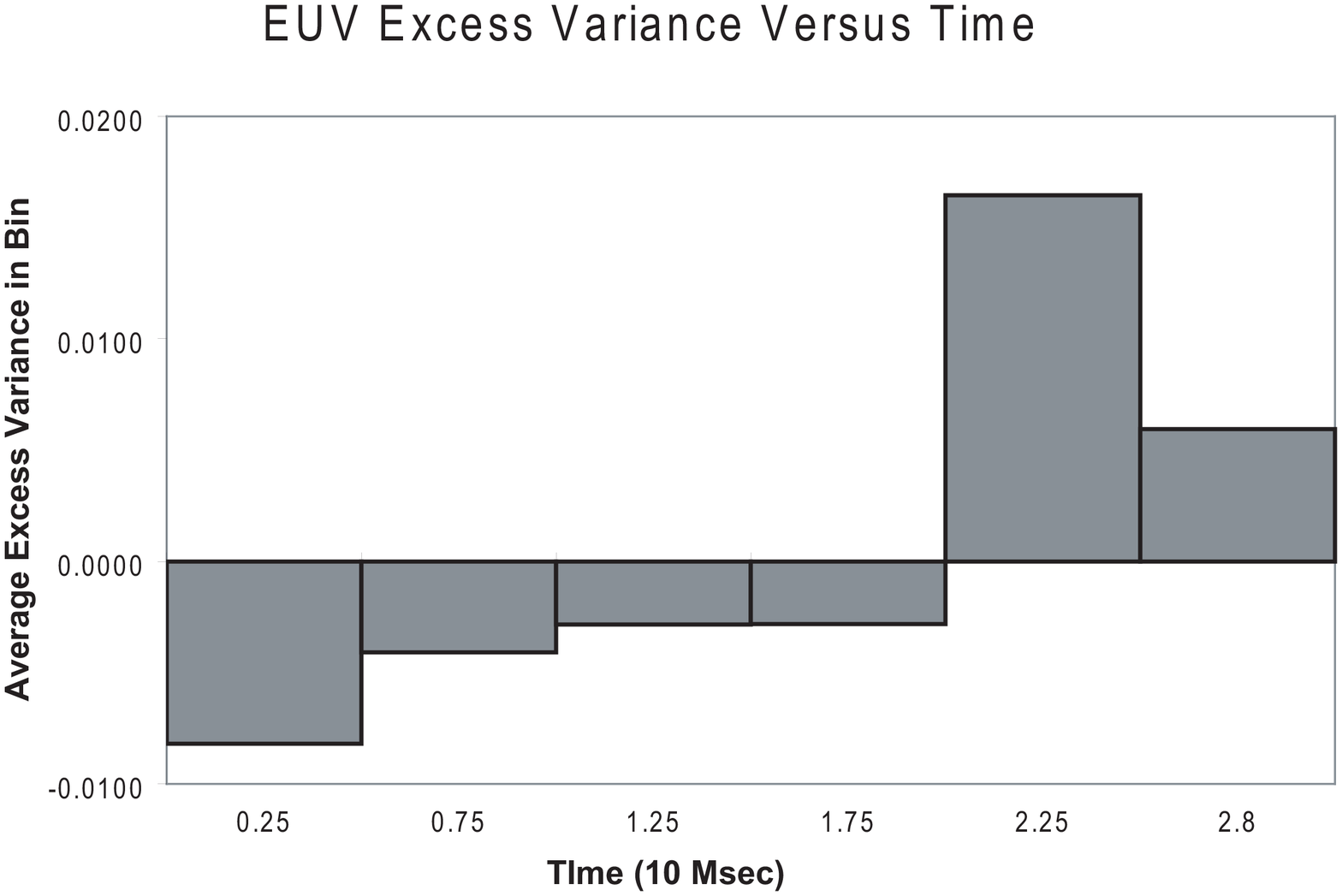}
\includegraphics[width=80 mm, angle= 0]{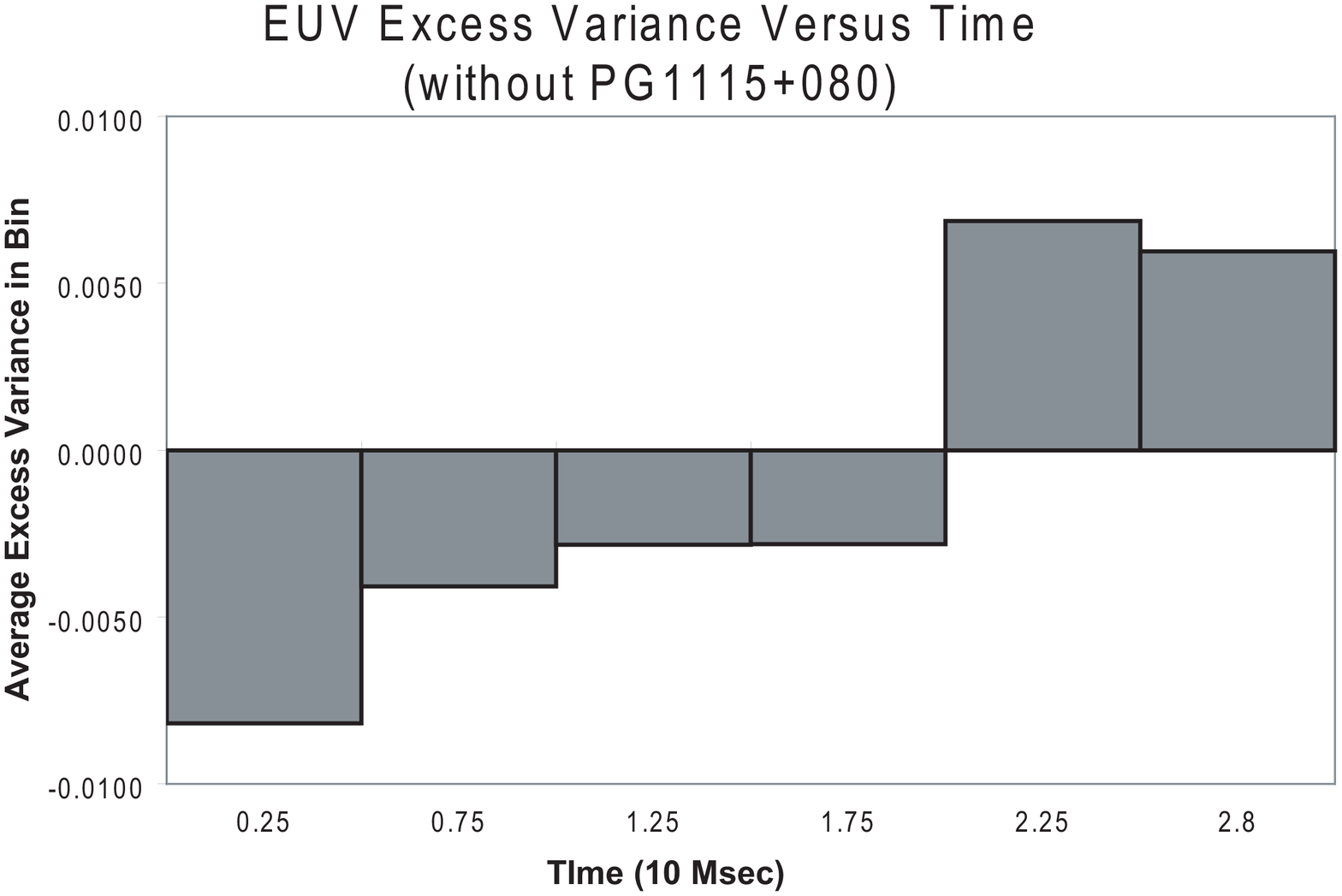}
\caption{The left hand frame is a histogram of the excess variance
versus time between observations. The most straightforward
interpretation is that the EUV continuum of the ensemble of quasars
shows no variation beyond that expected by random chance for time
intervals $<2.0 \times 10^{7}$s and the variability of the ensemble
of quasars far exceeds that expected by random chance for time
intervals $>2.0 \times 10^{7}$s to $3.1 \times 10^{7}$s. The right
hand frame is the same histogram, but with the extremely variable
quasar, PG 1115+080, removed. The result is qualitatively unchanged.
Statistically significant relative variation of the EUV continuum
occurs at $>2.0 \times 10^{7}$s at a typical value of 25\%. The
level of variability in the ensemble of quasars is approximately
constant up to $3.1 \times 10^{7}$s.}
\end{center}
\end{figure}
\subsection{The Affects of the Minimum Flux Cutoff}
In this section, we discuss the ramifications of altering the SDSS
minimum flux cutoff. If chosen properly, this should not affect the
results significantly, but should exclude some outliers with a
variability enhanced by the random noise level. To start with, the
left hand side of Figure 3 is a histogram of the distribution of
variation as a function of time using the data in the left hand
frame of Figure 2. Each bar of the histogram represents the average
variation of the individual quasars in that bin of time (each
individual quasar variation is computed per Equation (4)). There are
two items that can be explored with this depiction of the data.
Firstly, there is clearly much more variability for time separations
$> 2 \times 10^{7}$ s compared to time separations $< 1.5 \times
10^{7}$ s. The variability is larger at $>99.9\%$ level of
statistical significance based on a Kolmogorov-Smirnov test. The
other issue that is explored in this frame as well as the right hand
frame of Figure 3, is the robustness of our methods and the
threshold of detectable variability. Recall our efforts in the last
section to quantify the accuracy of the SDSS flux calibrations using
the technique of using adjacent field galaxies as an accurate
empirical secondary calibration. Since most of the sample are SDSS
quasars, the secondary flux calibration analysis indicates that bins
in the histogram with a variability $\sim 0.1$ are consistent with
no detected variability (just systematic flux measurement errors).
The implication of this is that (possibly with the exception of a
few mildly variable outliers) there is no measurable EUV variation
for separation times $< 1.5 \times 10^{7}$ s. More precisely since
the range between $1.0 \times 10^{7}$ s and $1.5 \times 10^{7}$ s is
sparsely sampled (see the left hand frame of Figure 2), a more
rigorous statement supported by the data is, there is no measurable
EUV variation (possibly with the exception of a few mildly variable
outliers) for separation times $< 1.0 \times 10^{7}$ s. Yet,
significant variation is common for separation times $> 2 \times
10^{7}$.
\par The right hand frame of Figure 3 is used to explore the issue
of the average flux density lower cutoff for the SDSS sample, $3.0
\times 10^{-17} \rm{erg/s}/\rm{cm}^{2}/\AA$. One issue with the
cutoff is that it would automatically reject large variation for the
fainter objects. Thus, highly variable sources are underestimated is
general in all bins in the left hand frame of Figure 3. For the bins
on the right hand side of the histogram with significant average
variation, adding a couple more sources with a modest or large
variation will not modify the average noticeably. However, adding
one highly variable source to the bins on the left can modify the
average variation proportionately much more than in the bins on the
right. In order to explore this we looked at changing the average
flux density lower cutoff for the SDSS sample to $2.0 \times
10^{-17} \rm{erg/s}/\rm{cm}^{2}/\AA$. This increased the number of
epochs from 54 to 64 in the SDSS sample. The histogram for the lower
cutoff is shown in the right hand frame of Figure 3. The general
trend is the same as Figure 2. However, one highly variable source
in the second bin raised this average considerably. Even so, there
is clearly much more variability for time separations $> 2 \times
10^{7}$ s compared to time separations $< 1.5 \times 10^{7}$ s. The
variability is still larger at $>99.9\%$ level of statistical
significance based on a Kolmogorov-Smirnov test.

\subsection{Excess Variance Analysis} In order to accurately assess variability one must
incorporate the uncertainty of the measurements, $\sigma_{i}$, into
the measure of variability. In this section, we use the excess
variance in order to segregate the role of the uncertainty (as
determined above to be mainly systematic in nature due to our sample
selection criteria) in our measurements of the intrinsic
variability. As discussed above, in regards to the background field
galaxy analysis, the systematic uncertainty in the absolute flux
measurements of the SDSS sample in Table 3 is $\approx10\%$, or
$\sigma_{1} \approx 0.1F_{1}$ and $\sigma_{2} \approx 0.1F_{2.}$.
Our detailed discussion of the selection criteria for the HST data
found a similar systematic uncertainty. Thus, on a statistical
basis, one can only detect relative variability larger than
$\sqrt{\sigma_{1}^{2} +\sigma_{2}^{2}}$. This concept is typically
expressed as the excess variance, $\sigma_{\rm{rms}}^{2}$,
\citep{nan97}

\begin{equation}
\sigma_{\rm{rms}}^{2}=\frac{1}{N\mu^{2}}\sum_{i=1}^{N}
[(F_{i}-\mu)^2- \sigma_{i}^{2}]\; ,
\end{equation}
where $\mu$ is the average of the flux measurements, $F_{i}$. This
is an odd application of excess variance considering that we only
have two values of "i". However, it precisely describes the
circumstance of interest: what constitutes variation larger than
that expected by random chance? We note that we can express
$\sigma_{\rm{rms}}^{2}$ in terms of the relative variability, $V$,
in Tables 2 and 3 using Equation (4),
\begin{equation}
\sigma_{\rm{rms}}^{2}=\left(\frac{V}{2+V}\right)^{2}-
\frac{0.02}{(2+V)^{2}}[1+(1+V)^{2}]
\end{equation}
Figure 4 are scatter plots of $\sigma_{\rm{rms}}^{2}$ versus the
time between observations. On the left hand side, the time is
measured in seconds and on the right hand side, the time is measured
in estimated geometrized units of the black hole mass. The plots are
similar to those of Figure 2. However, the uncertainty associated
with each value of relative variability, $V$, in Figure 2 is
extricated by the defining relationship for $\sigma_{\rm{rms}}^{2}$
in Equation (8). The minimum criteria for a variability detection
consistent with the uncertainty in each individual flux measurement
is $\sigma_{\rm{rms}}^{2} > 0$. With this criteria for variability,
4/40 of the quasar observation pairs with a time separation $<
2\times 10^{7}$ sec are variable and 21/38 of of the quasar
observation pairs with a time separation $> 2\times 10^{7}$ are
variable.  Figure 5 is a histogram highlighting the almost switch-on
like behavior of the variability once the statistical uncertainty of
each measurement is accounted for in the definition of variability.
We note that $\sigma_{\rm{rms}}^{2}$ averaged over a bin is similar
to the structure function used in the UV photometry studies of
quasars \citep{wel11,mac12}. However, there is not much evidence in
the excess variance analysis of the EUV continuum variability of a
gradual increase of variability for small time intervals to larger
time intervals that was seen for the UV photometric structure
functions. We eliminate the most variable source from the study in
the right hand frame of Figure 5 in order to see if this is skewing
the analysis. It does not. There is still more of a switch-on
behavior than a gradual turn on of variability. We explore this in
more detail in the remainder of the subsection.
\begin{table}
\caption{Probability Matrix for Excess Variance\tablenotemark{a}}
 \begin{tabular}{ccccc}
\hline
   Bin & 2.0 - 3.1 $10^{7}$ s & 2.0 - 2.5 $10^{7}$ s & 1.0 - 2.0 $10^{7}$ s \\
\hline
 2.5 - 3.1 $10^{7}$ s & N/A & 0.328  & 0.252\\
 1.0 - 2.0 $10^{7}$ s & 0.101  &  0.083   & N/A  \\
 0.5 - 2.0 $10^{7}$ s & 0.046  &  0.057   & N/A  \\
 0.0 - 2.0 $10^{7}$ s & $<0.001$ & 0.003 &  N/A\\
 0.0 - 1.5 $10^{7}$ s & $<0.001$ & 0.003 &  N/A\\
 0.0 - 1.0 $10^{7}$ s & $<0.001$ & 0.002 & 0.051 \\
 0.0 - 0.5 $10^{7}$ s & $<0.001$ & 0.002 & $0.004$  \\
 \hline
\end{tabular}
\tablenotetext{a}{The entries are the Kolmogorov-Smirnov
probabilities that the data in the bins are drawn from the same
sample}
\end{table}
\par We can characterize some of the behavior of
$\sigma_{\rm{rms}}^{2}$ denoted in Figures 4 and 5 by computing a
probability matrix based on the Kolmogorov-Smirnov test. The results
are given in Table 4. The matrix is a collection of
Kolmogorov-Smirnov tests. Each entry compares two bins with at least
10 pairs of observations, The value that is recorded is the
probability that the excess variance in the two bins are drawn from
the same population. There are four statistical inferences that can
be drawn from the probability matrix,

\begin{enumerate}
\item $\sigma_{\rm{rms}}^{2}$ is larger when the time between observations in the quasar rest frame is $>2.0 \times
10^{7}$s then it is when the time between observations in the quasar
rest frame is $<1.0 \times 10^{7}$s at $>99.9\%$ statistical
significance.

\item There is no statistical significance difference in
$\sigma_{\rm{rms}}^{2}$ whether the time between observations in the
quasar rest frame is $2.0 -2.5 \times 10^{7}$s or if the time
between observations in the quasar rest frame is $2.5-3.1 \times
10^{7}$s.

\item There is no statistical significance difference in
$\sigma_{\rm{rms}}^{2}$ whether the time between observations in the
quasar rest frame is $1.0 -2.0 \times 10^{7}$s or if the time
between observations in the quasar rest frame is $>2 \times
10^{7}$s.

\item There is a marginal statistical difference in $\sigma_{\rm{rms}}^{2}$ when the time between observations in the quasar rest
frame is $0.0 -1.0 \times 10^{7}$s as compared to the time between
observations in the quasar rest frame is $1.0- 2.0 \times 10^{7}$.

\end{enumerate}
Consider these points in the context of Figure 5.  Based on points 3
and 4 and Figure 5, there is an increased level of
$\sigma_{\rm{rms}}^{2}$ above $1.0 \times 10^{7}$s, but
$\sigma_{\rm{rms}}^{2}<0$ for time intervals $<2.0 \times 10^{7}$s
indicating that this increase is fairly mild and the variability is
not statistically significant compared to the level of the
measurement uncertainty. From point 1 and the Figure 5 for time
intervals $>2.0\times 10^{7}$s, $\sigma_{\rm{rms}}^{2}$ has
increased to a level that far exceeds the threshold for
statistically significant variability for many quasars. Yet, by
point 2 and Figure 5, the propensity and magnitude of variability
does not increase for time intervals $> 2.5 \times 10^{7}$s. This
last point does not support a monotonic increase in
$\sigma_{\rm{rms}}^{2}$ as the time interval increases unless the
rate of increase diminishes drastically for time intervals $> 2.5
\times 10^{7}$s. The overall behavior is well described by a broad
threshold for measurable variability that occurs typically in the
range of time intervals between observations of $1.5 -2.0 \times
10^{7}$s above which the variability is roughly constant for time
intervals less than 1 year. This behavior is similar to what has
been seen in UV and optical quasar structure functions that are
based on broad band photometry \citep{wel11,mac12}. The photometric
structure functions show a steady increase in variability as the
time between observations increases to a few hundred days. Beyond
this time interval, the frequency of variability is roughly
constant. The one difference is that the EUV continuum variability
shows signs of increasing more abruptly after 100 days rather than
an extrapolation of a steady increase from the small time intervals
as in the photometric UV structure functions. This might be a result
of the small number statistics in this study in the range $1.5 -2.0
\times 10^{7}$s, or a fundamental difference associated with the
fact that the EUV region is at the inner edge of the accretion disk.
We note that a structure function cannot be used on the data
presented here since we do not eliminate low variability sources as
in photometric structure functions \citep{wel11}. Consequently, the
variability can be less than the estimated error (see the negative
values of excess variance in Figures 4 and 5) rendering a putative
structure function at time intervals, $<2.0\times 10^{7}$s, to
values that are imaginary numbers (i.e., basically the square root
of the negative excess variance).
\par In summary, the excess variance analysis supports a rapid
change in the likelihood of significant variability for time
intervals between observations that exceed some threshold above $
1.5 \times 10^{7}$s. The threshold is loosely constrained due to
small number statistics in the crucial range of $1.5 - 2.0 \times
10^{7}$s. The likelihood of variability levels off above above $ 2.5
\times 10^{7}$s after crossing this poorly resolved broad threshold.
The threshold is a broad region because it clearly depends on many
properties such as black hole mass and accretion rate. For this
extreme luminosity quasar sample, virtually all the black hole
masses are large and all the accretion rates very high making the
threshold region somewhat concentrated in time and this enhances its
appearance in Figure 5.
\begin{figure}
\begin{center}
\includegraphics[width=125 mm, angle= 0]{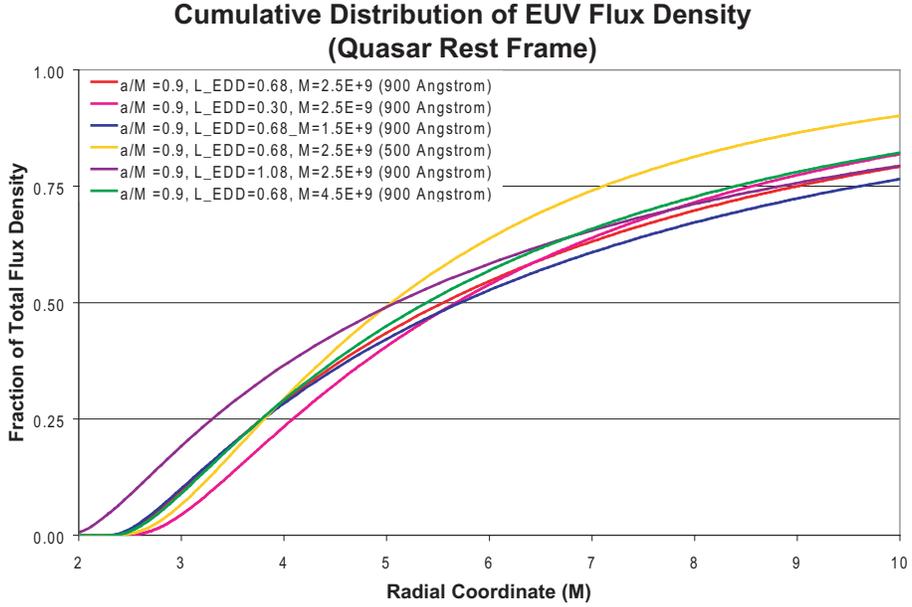}
\caption{The cumulative distribution of EUV continuum flux density
in slim disk models. The spin that is chosen is the fiducial spin
motivated in Section 5 based on theory and empirical study. The
black hole masses and Eddington rates are typical of those estimated
for the quasars in Tables 2 and 3. The y-axis indicates how much of
the total EUV flux density (at the designated wavelength) is emitted
from the disk surface as a function of radius. A value of 1.0 is
100\% of the flux density and 0.5 is 50\% of the total flux density.
Notice that the majority of the EUV flux at all wavelengths is
emitted inside of $ r = 7M $ for all models. According to these slim
disk models the variability in the EUV continuum is a result of the
local physics at $ r < 6M - 8M $.}
\end{center}
\end{figure}
\begin{figure}
\begin{center}
\includegraphics[width=115 mm, angle= 0]{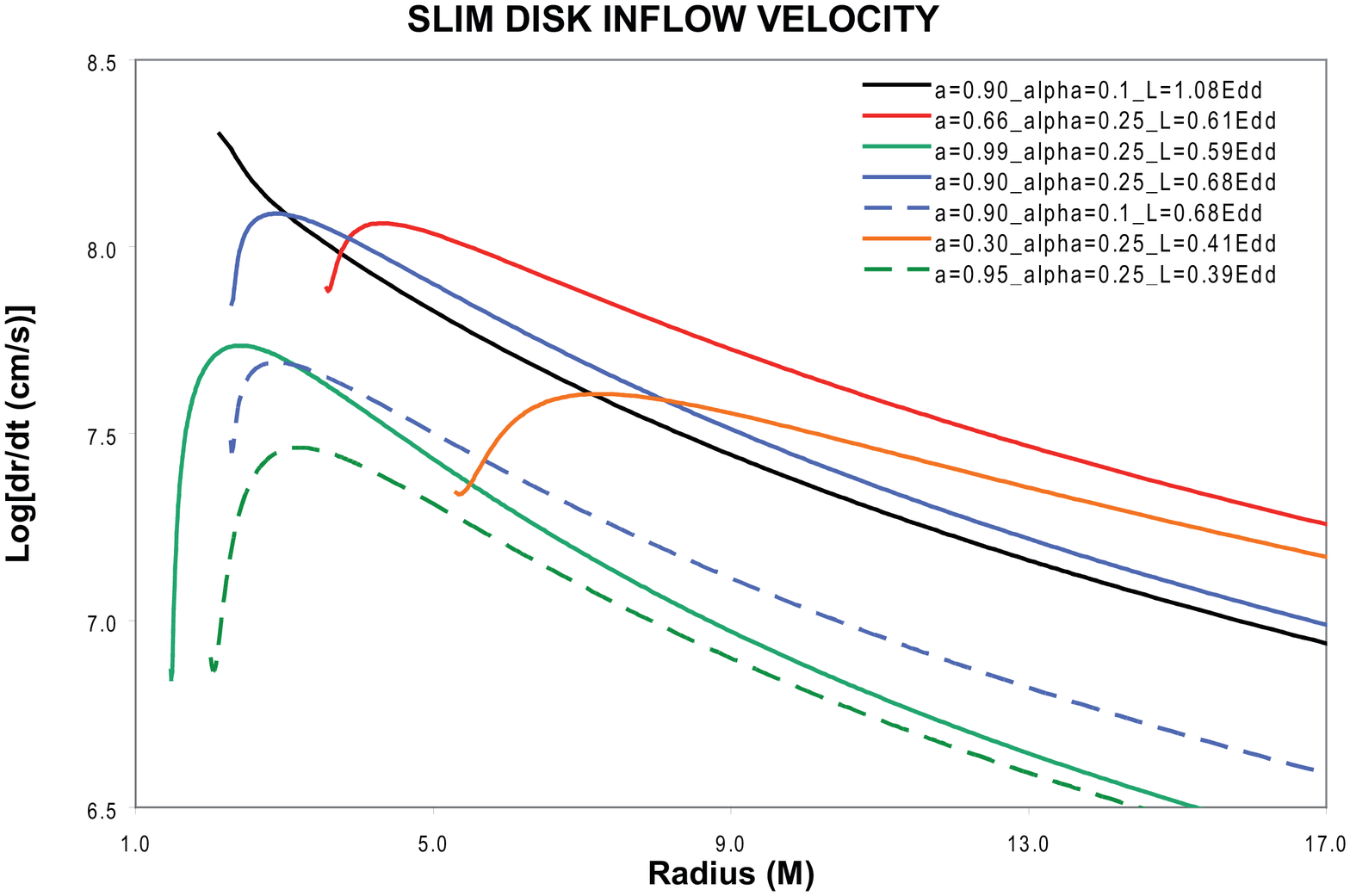}
\caption{The inflow velocity, $|dr/dt|$, for various slim disk
models. The inflow velocity is vertically averaged over the height
of the slim disk. The extremely large value of $\alpha = 0.25$ is
plotted as an upper bound on the magnitude of the infall velocity of
the optically thin surface layers of the disk, The inflow velocity
is orders of magnitude less than the speed of light.}
\end{center}
\end{figure}
\section{Slim Disk Models of EUV Variation} The previous section
dealt with an empirical description of the data. This section is an
attempt to make a speculative assessment on the implied size of the
EUV emitting region based on models of slim disks \citep{sad11}.
Being located just shortward of the peak of the SED, the EUV
continuum that is sampled in Tables 1 - 3 likely represents the
highest frequency optically thick thermal emission from the
accretion flow. This emission therefore comes from the innermost
regions of the accretion disk. We do not hypothesize on the
nontrivial radiative transfer of this emission through an
atmosphere. This includes, amongst other possibilities, electron
scattering, \citet{ear78,cze87}, and absorption that drives an
outgoing wind \citep{lao14}. We just assume that a thermal EUV
component was emitted from the disk, but acknowledge that its
amplitude might be altered by processes in an enveloping atmosphere.

From Tables 2 and 3, we expect that the Eddington rate would be too
large for the standard thin disk model to be accurate
\citep{sha73,nov73}. For $R_{\rm{Edd}} >0.2$, several assumptions of
thin disk theory and the approximations associated with the vertical
integration of slim disks start to break down \citep{szu96}. Thus,
we consider the most recent slim disk model solutions \citep{sad11}.
Empirically, there is a change in several spectral properties at
$R_{\rm{Edd}} >0.2$ that provides observational evidence to an
accretion mode change at $R_{\rm{Edd}} \approx 0.2$. This has been
characterized by the designation of Population A ($R_{\rm{Edd}}
>0.2$) and Population B quasars \citep{mar04}. Our sample is
entirely Population A type spectra. The classic argument of
\citet{bar70} indicates that black holes should be spun up near
their maximum allowed value if subject to prolonged accretion.
Prolonged accretion seems reasonable for luminous quasars
considering the He II proximity effect, high redshift quasars can
accrete at a high rate for $>34$ Myr \citep{syp14}. The
\citet{bar70} conclusion has been corroborated empirically by
considering the X-ray background, which requires high efficiency
accretion associated with near maximal spin \citep{elv02}. The
assumption of the Bardeen argument is that the rotation of the
galaxy determines a preferred sign of angular momentum for the
accreted matter. We acknowledge that there might be cases in which a
complicated accretion history might not lead to a large net spin up
(i.e., possible flips in the sign of angular momentum associated
with merger scenarios), but we do not consider this typical.
Consequently, based on the Bardeen calculation and the He II
proximity effect, we assume that most of our quasar sample have a
high spin rate. We still consider low spin in the following, but
high spin is the nominal configuration. The axisymmetric, time
stationary spacetime (Kerr-Newman) metric is uniquely determined by
three quantities, $M$, $a$ and $Q$, the mass, angular momentum per
unit mass, and the electric charge of the black hole respectively.
For the sake of this paper, we consider $Q$ negligibly small. In
Boyer-Lindquist coordinates $(t,\, r,\, \phi,\, \theta)$, the event
horizon, $r_{{+}}$, is
\begin{equation}
r_{{+}}=M+\sqrt{M^{2}-a^{2}} \; .
\end{equation}
 Based on the Bardeen calculation and the He II proximity effect, we assume that most of our
quasars are near maximal rotation rate, $a/M \sim 1$. In particular,
we will only discuss accretion disks with $0.99\geq a/M \geq 0.3$.
As mentioned above, there can be individual quasars in which the
accretion disk angular momentum vector changed directions over time,
but these are considered outliers that will not affect the overall
trends of the sample. Our fiducial value of spin is conservatively
modest in our opinion, $a/M =0.9$.
\subsection{The Location and Inflow Velocity of the EUV Emitting
Gas}
\par The first thing to consider is the location of the optically
thin surface layer that is the source of the EUV continuum emanating
from the appropriate slim disk models for the Eddington rates and
masses that are estimated in Tables 2 and 3. Figure 6 plots the
expected location of the EUV emitting plasma for various slim disk
models. The range of masses are typical of the luminous quasars that
comprise the samples in Tables 2 and 3 and so are the ranges of
Eddington luminosity. We choose a typical line of sight for a quasar
that is $30^{\circ}$ from the accretion disk normal
\citep{ant93,bar89}. In the following, changing the line of sight to
$0^{\circ}$ makes small changes that do not affect any of the
conclusions. The plot is a cumulative distribution of the flux
density at $900\AA$ and $500\AA$ that is emitted from the disk. For
example, a value of 0.5 means that half of the observable flux
density is emitted inside this radius. The flux density is computed
in the cosmological rest frame of the host galaxy. Thus, it includes
the Doppler shifting (including both the transverse Doppler shift
and the gravitational Doppler shift) of the disk emission. The disks
are characterized by a vertically averaged viscosity parameter of
$\alpha= 0.1$. The emitted flux distribution from the slim disk is
very weakly dependent on the value of $\alpha$ (A. Sadowski private
communication 2016). For all the models in Figure 6, 50\% of the
emission comes from inside a radius that is between 5M and 6M. For
higher black holes spins these curves shift to the left and the 50\%
point shifts that way as well. As an extreme example, if $a/M=0.99$
and $R_{\rm{Edd}}=0.20$, the 50\% point is at 3.8 M for the $900\AA$
flux density. If an extreme change happens to the emissivity of the
optically thin surface layer that is the source of 50\% - 60\% of
the the $900\AA$ flux density, this could clearly create the 25\% -
30\% variability that typifies the behavior of most of the variable
quasars in Tables 2 and 3 and Figures 2 - 5. Based on Figure 6, we
look for a variability time scale associated with the optically thin
surface layer at $r < 8M$ in the slim diks models that is on the
order of $1.5 - 2.5\times 10^{7}$ s or 1500 - 2500 M in geometrized
units based on Figures 2 - 5.
\par The primary parameter to consider is the infall velocity of the optically thin surface
layer. The infall time from a particular radius in the EUV portion
of the disk can be computed from the radial velocity in
Boyer-Lindquist coordinates, $dr/dt$. Figure 7 plots the radial
velocity for various relevant sets of parameters. The figure allows
for an exploration of the dependence of $dr/dt$ on the viscosity
parameter, the Eddington ratio and the black hole spin. One thing to
notice is the local maximum in $dr/dt$ at small radius and high
spin. This is a consequence of gravitational redshifting and the
centrifugal barrier. The Boyer-Lindquist coordinates are those of
observers that are stationary with respect to asymptotic infinity.
These observers never see any trajectory cross the event horizon.
The trajectories approach the event horizon asymptotically slowly,
the freezing of the flow \citep{pun08},
\begin{equation}
\lim_{r\rightarrow r_{{+}}} \frac{dr}{dt} =
-\frac{r_{{+}}^{2}-2Mr+a^{2}}{r_{{+}}^{2} +a^{2}} \propto (r -
r_{{+}}) \rightarrow 0\;.
\end{equation}
When the spin increases, the inner edge of the disk approaches the
event horizon. Thus, for high spin, the gravitational redshift
effects will be large near the inner edge of the disk and in the
plunge region (the region where gravitational attraction towards the
black hole overwhelms centrifugal force). Since the infalling gas
never actually reaches the event horizon in Boyer-Lindquist
coordinates, the time for infall is not well defined. Furthermore,
the solutions of \citet{sad11} do not extend all the way to the
event horizon. These issues are circumvented by the condition that
infall is measured to the r coordinate that is halfway between
$r_{{+}}$ and the inner edge of the disk. We note two points:
\begin {itemize}
\item In general, the time for the gas to traverse this gap from the
inner edge of the disk to this midpoint is at most of few percent of
the infall times that we compute.
\item Furthermore, due to gravitational redshifting
the flux detected by observers at asymptotic infinity is negligible
for emission between this midpoint and the event horizon.
\end{itemize}

For these two reasons, this simple definition of the infall end
point leads to very robust results that are not sensitive to precise
location of the end point. The primary result of Figure 7 is that
the magnitude of $dr/dt$ within the accretion disk is orders of
magnitude less than the speed of light. Thus the infall times in
geometrized units are orders of magnitude larger than the radius
from which the infall begins! We also note the scalings displayed in
Figure 7, indicate that $|dr/dt|$ increases (decreases) with
increases (decreases) of $\alpha$ and $R_{\rm{Edd}}$ and decreases
(increases) of $a/M$.
\begin{figure}
\begin{center}
\includegraphics[width=125 mm, angle= 0]{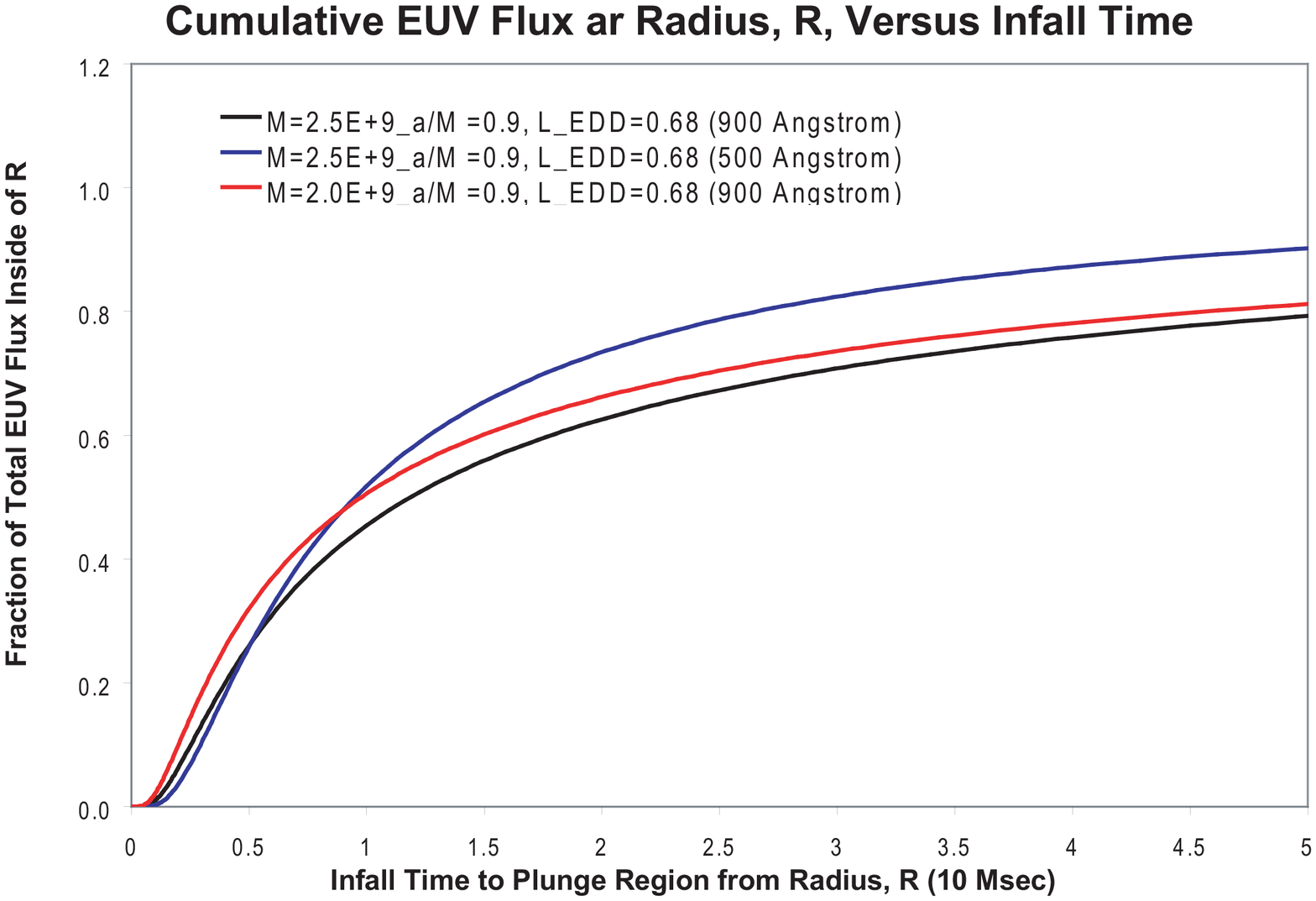}
\includegraphics[width=125 mm, angle= 0]{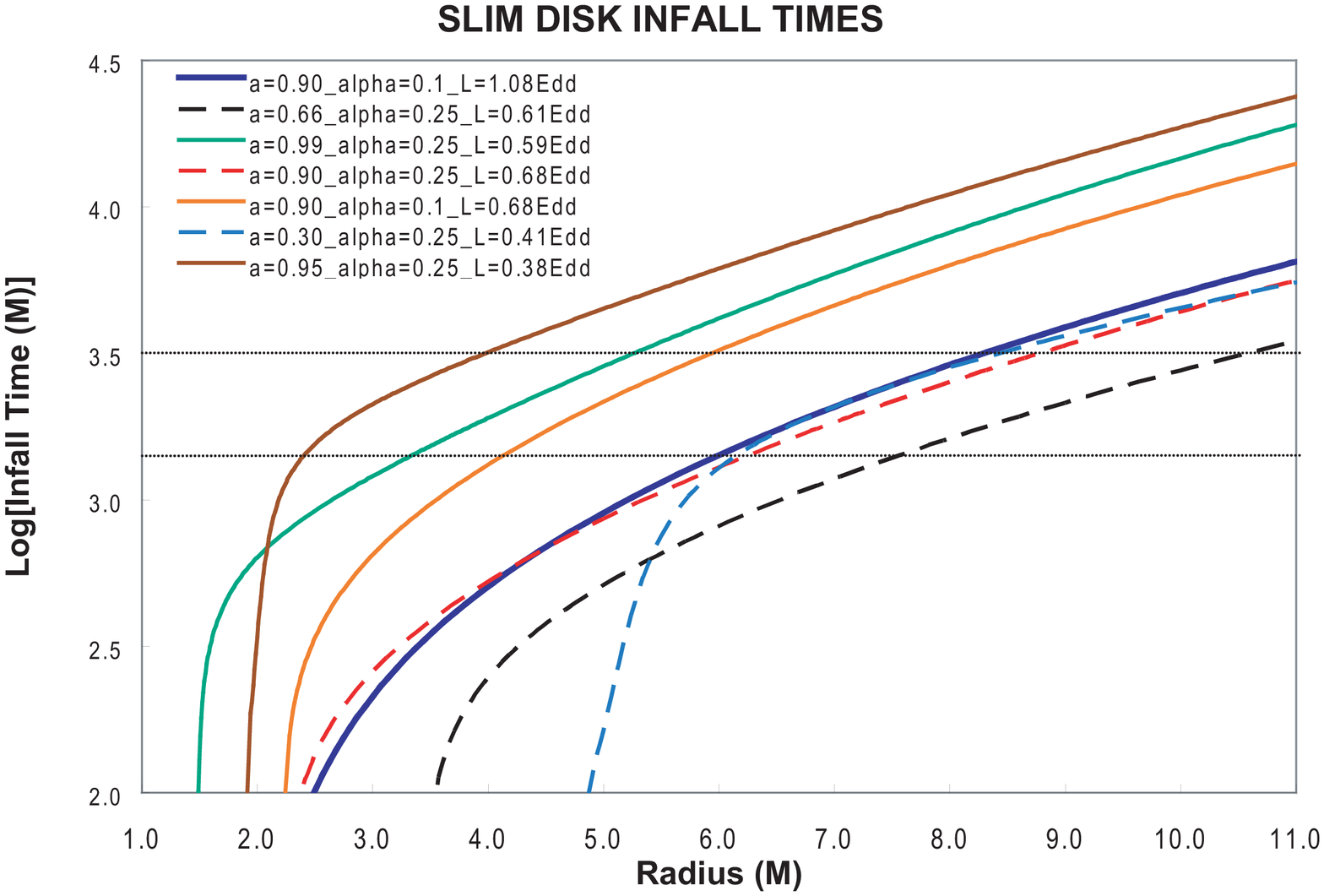}
\caption{The top frame shows a plot of the fraction of the
cumulative $900 \AA$ flux density inside of a radius ,R,
$F_{\lambda}(\lambda =900\AA)\mid_{r<R}$ versus the infall time from
a radius, R, $T_{\rm{infall}}(R)$ based on the radial velocities
that are calculated as in Figure 7. This indicates that the infall
time from the EUV emitting region is a viable source of the time
scale for EUV variability The bottom frame plots the infall time
based on slim disk models for a variety of black hole parameters. It
plots the infall time (to a point mid way between the event horizon
and the inner edge of the accretion disk) as a function of start
radius for the trajectory. This is computed for the same slim disk
parameters as were used in Figure 7. The horizontal dashed lines
correspond to time intervals associated with variability that were
shown in the right hand side of Figure 2.}
\end{center}
\end{figure}
\par Notice that most of the curves in Figure 7 are based on an
extremely large value of the viscosity parameter, $\alpha = 0.25$.
The motivation for this is the following. The slim disk models are
very optically thick except for a surface layer of optically thin
gas. This is the gas that is responsible for the observed EUV flux.
The slim disk has vertical structure and the velocity that is
calculated in \citet{sad11} is vertically averaged. We are actually
only interested in the radial velocity of the optically thin surface
layer. This is beyond slim disk models and requires full 3-D
calculations. Although the vertical structure of slim disks is not
well known, recent 3-D numerical simulations indicate that near the
inner edge of the disk, surface layers move inward approximately
twice as fast as the vertical average \citep{sad16}. This can be
represented by a higher effective $\alpha$ in the surface layers
that is twice as large as the vertical average (A. Sadowski, private
communication 2016). Hence, the focus on $\alpha = 0.25$ in the
Figure 7.
\par The first thing to consider is the implication of Figure 7 for the infall times for the surface
layers of the slim disk that are responsible for the EUV continuum
in Figure 6.  The cumulative $900 \AA$ flux density in Figure 6 and
the infall times in bottom frame of Figure 8 are both functions of
the radial coordinate, $R$. We exploit this in the top frame of
Figure 8 by plotting these implicit functions of $R$, without any
direct reference to $R$. The top frame of Figure 8 is a plot of the
fraction of the cumulative $900 \AA$ flux density inside of a
radius, R, $F_{\lambda}(\lambda =900\AA)\mid_{r<R}$ versus the
infall time from a radius, R, $T_{\rm{infall}}(R)$ based on the
radial velocities that are calculated as in Figure 7. For typical
values of black hole mass and Eddington ratio
$T_{\rm{infall}}(R)\mid_{R=6M-7M}\approx 2.0 \times 10^{7}$s,
$F_{\lambda}(\lambda =900\AA)\mid_{r<7M}$ is 50\% to 65\% of the
total $F_{\lambda}(\lambda =900\AA)$ emitted from the disk. Large
changes to the surface layer in this region will affect a sufficient
fraction of the EUV emitting gas in order to account for the $\sim
25\% - 30\%$ variability in Tables 2 and 3 as well as Figures 2 - 5.
Thus, we have established that the infall time associated with the
EUV emitting surface layer of the appropriate slim disk model is
consistent with the variability time scale indicated in Figure 5.
The bottom frame of Figure 8 uses the $|dr/dt|$ values from Figure 7
to integrate trajectories and determine the infall times for
different starting points. The infall time in geometrized units is
plotted versus the radial coordinate of the initiation point of the
infall for various designated slim disk models. The plot indicates
that the infall time decreases (increases) with increases
(decreases) of $\alpha$ and $R_{\rm{Edd}}$ and decreases (increases)
of $a/M$. The horizontal dotted lines represent the range of infall
times in geometrized units for the significantly variable sources in
our combined sample of quasars in the right hand frame of Figure 1.
\par Figure 9 is more ambitious. It is a plot of the infall radius,
$R_{\rm{infall}}$, for the quasars in Tables 2 and 3 based on the
estimated mass, and spin of the black hole, versus the variability.
These values were computed from the slim disk models and were
tabulated in column (3) of Table 2 and column (12) of Table 3. On
the top of the figure is a plot of the relative variability versus
the maximum infall radius, $R_{\rm{infall}}$. The maximum infall
radius was obtained by associating the infall time from
$R_{\rm{infall}}$ (as depicted in bottom frame of Figure 8) with the
interval between observations. The calculation assumes the fiducial
model of $a/M=0.9$. The bottom plot in Figure 9 is a plot of the
excess variance as a function of $R_{\rm{infall}}$. The advantage of
these plots is that they remove the dependence of black hole mass
and Eddington rate. The disadvantage is that the data is randomized
by the significant systematic uncertainty in the black hole mass
estimates. However, as an ensemble of data, one expects that the
general trends should be valid. The scatter plots indicate that in
general, significant variability is seen for infall radii,
$R_{\rm{infall}}> 7 M$. This location is consistent with the
distribution of EUV flux depicted in Figure 6.
\par There is a curious extreme point in the scatter plots in Figure 9.
Consider the value of 5.5 M for the maximum infall radius of PG
1115+080. This has a natural explanation within the slim disk
models. Recall from Table 3 the extremely large black hole mass
estimate for this object $M_{bh} = 4.35 \times 10^{9} M_{\odot}$.
This is a cooler disk than our nominal disk with $M_{bh} = 2.5
\times 10^{9} M_{\odot}$. This shifts the $900\AA$ 50\% cumulative
distribution point down to 5M in Figure 6. We also see from Figure
6, that the $500\AA$ 50\% point is located at smaller values of
radius than the the $900\AA$ 50\% cumulative distribution point.
This is probably more relevant considering the spectrum of PG
1115+080 in Figure 1.
\subsection{A Crude Model of EUV Variability}
\par The discussions associated with Figures 6 - 9 indicate that
there is a possible link between plasma infall times from the EUV
emitting surface layer of the slim disk models and the EUV continuum
variability detected in Tables 2 and 3. With this motivation, we
consider the most straightforward explanation of EUV variability in
the context of slim disk models of quasars with the aid of the
schematic diagram in Figure 10. We acknowledge that this is a very
speculative and crude model. In the model, a significant change in
EUV luminosity is associated with an evacuation of the inner disk
plasma and a replenishment with new plasma with a different
emissivity. More specifically, since the equatorial regions of the
disk are optically thick, a significant change in EUV luminosity is
associated with an evacuation of the optically thin surface layer of
inner disk plasma and a replenishment with a new surface layer of
plasma with a different emissivity.
\begin{figure}
\begin{center}
\includegraphics[width=125 mm, angle= 0]{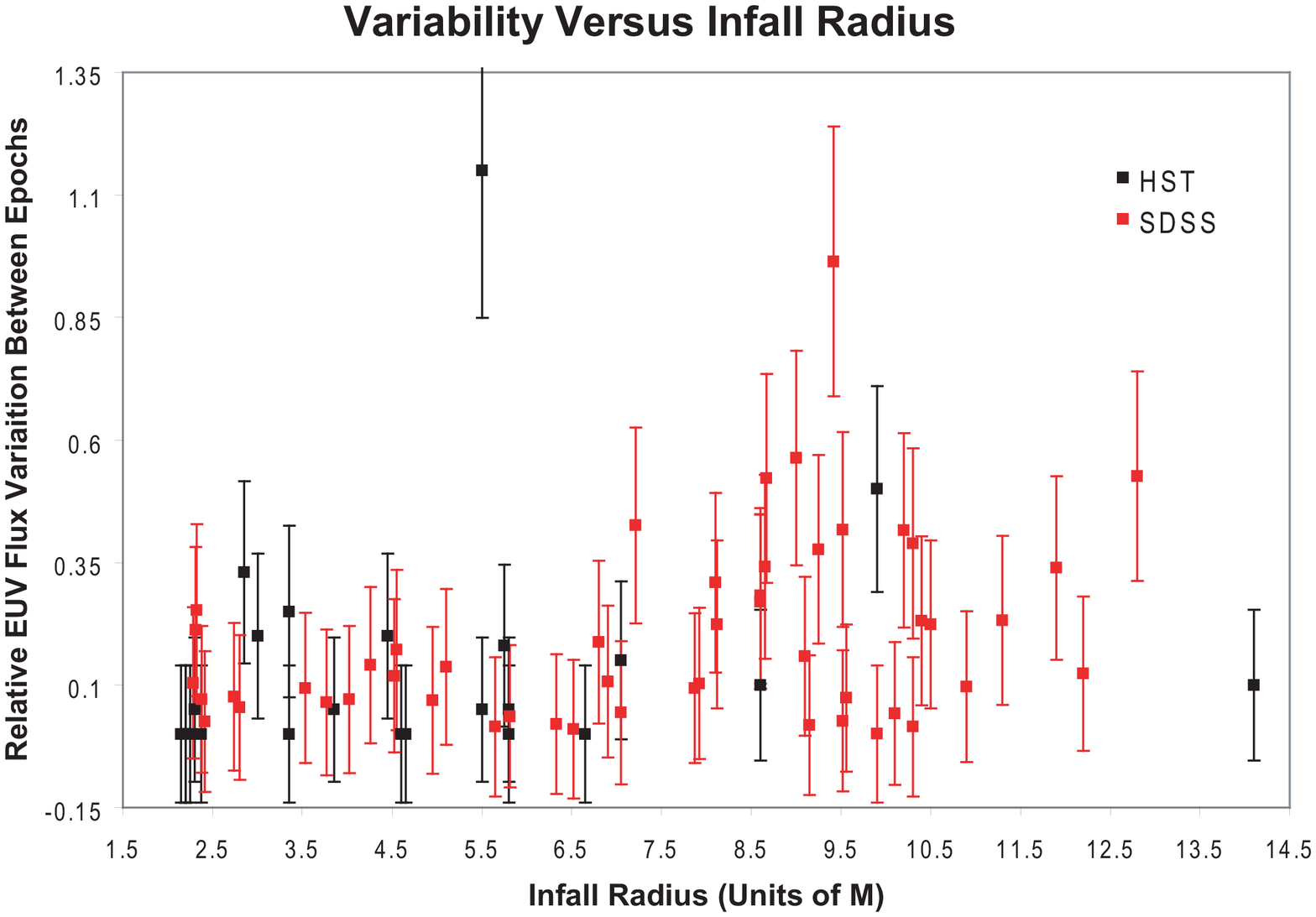}
\includegraphics[width=125 mm, angle= 0]{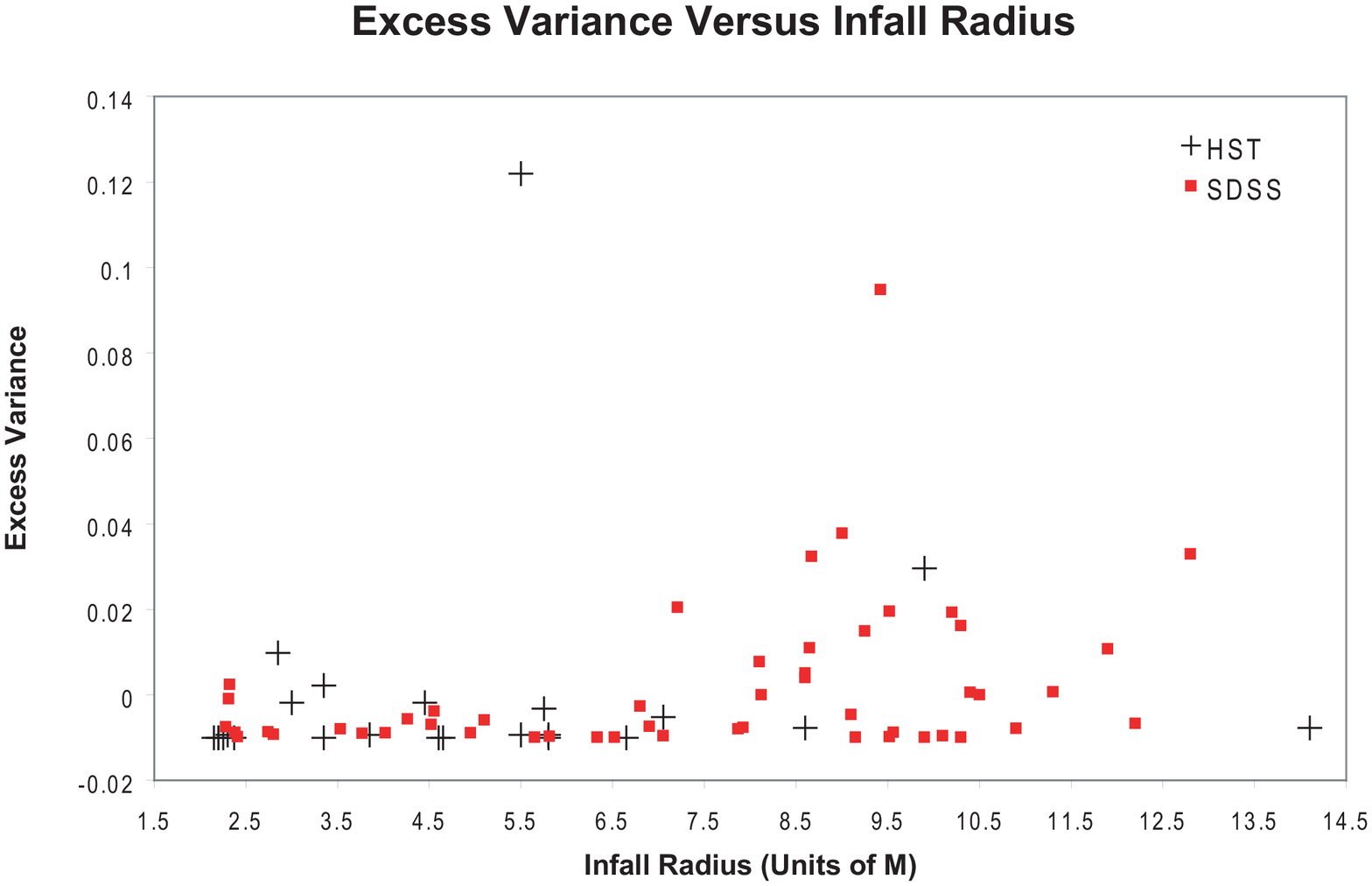}
\caption{The top frame plots the relative variability versus the
infall radius from Tables 2 and 3. This infall radius is computed
for the slim disk parameters that are given in Tables 2 and 3 by
equating the infall time of the optically thin surface layer at that
radius with the time between observations.. The bottom frame is a
plot of excess variance versus infall radius. These plots give
another crude measure of the location of the variable EUV emission.
We find that significant variability starts to occur at a value of
$r \approx 7M - 8.5M$ in both plots for our fiducial model,
consistent with the location of the EUV emitting region in Figures
6.}
\end{center}
\end{figure}

\par Figure 10 shows the basic configuration. The black hole is
depicted as a black disk. The optically thin surface layer of the
accretion disk is an annulus with an emissivity that is depicted in
grey scale, where black is the highest emissivity and white is zero
emissivity. In the schematic, there are two concentric circles that
demarcate the boundary of the EUV region. The boundaries can move
slightly as the accretion state changes, but this is a higher order
affect that is beyond this schematic representation. We describe the
particular circumstance depicted in each frame from left to right.
The frame on the left indicates an enhanced emissivity in dark
shading that envelopes the EUV region and extends beyond it. This
would be the configuration of a significant elevation of EUV flux in
magnitude and time. Within standard accretion disk theory, this
increase is associated with an increase of the effective temperature
of the surface layer \citep{sad11}. The second frame shows an epoch
with decreased emissivity. If we were to compare the two epochs, it
would correspond to some of the larger variations in our combined
sample of the last section. The last two frames indicate the
different types of EUV flares. A large variation in Table 3 $\sim
0.5$, would require that virtually all of the EUV region change
emissivity. For example, the third frame represents such a high
state that we consider an $O(1)$ variation. This is in distinction
to the last frame that has only a small region of the EUV annulus
filled with high emissivity gas and this might be represent a short
period of slightly elevated flux with a variation parameter $<0.2$.

\begin{figure}
\begin{center}
\includegraphics[width=80 mm, angle= 0]{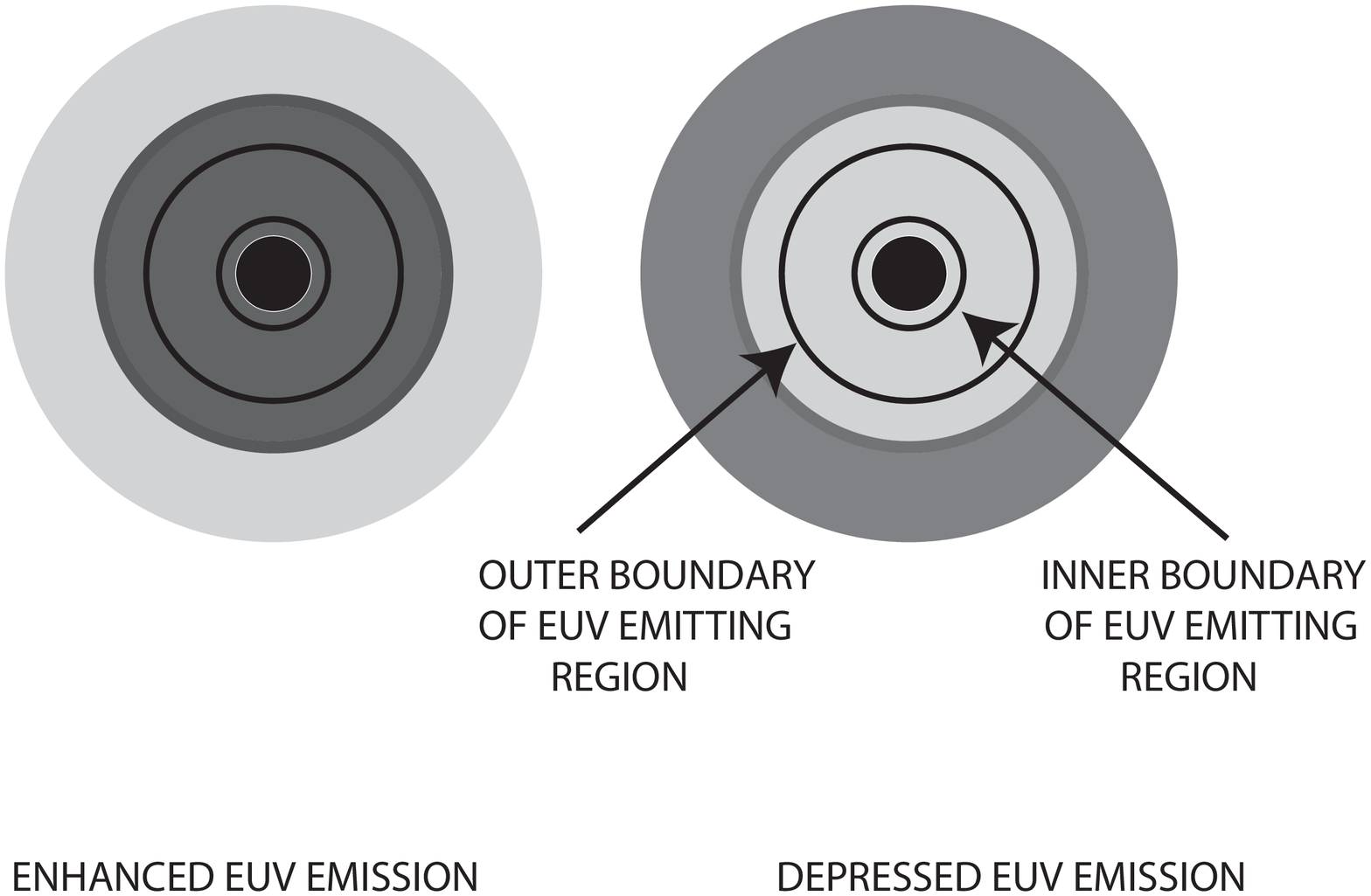}
\includegraphics[width=80 mm, angle= 0]{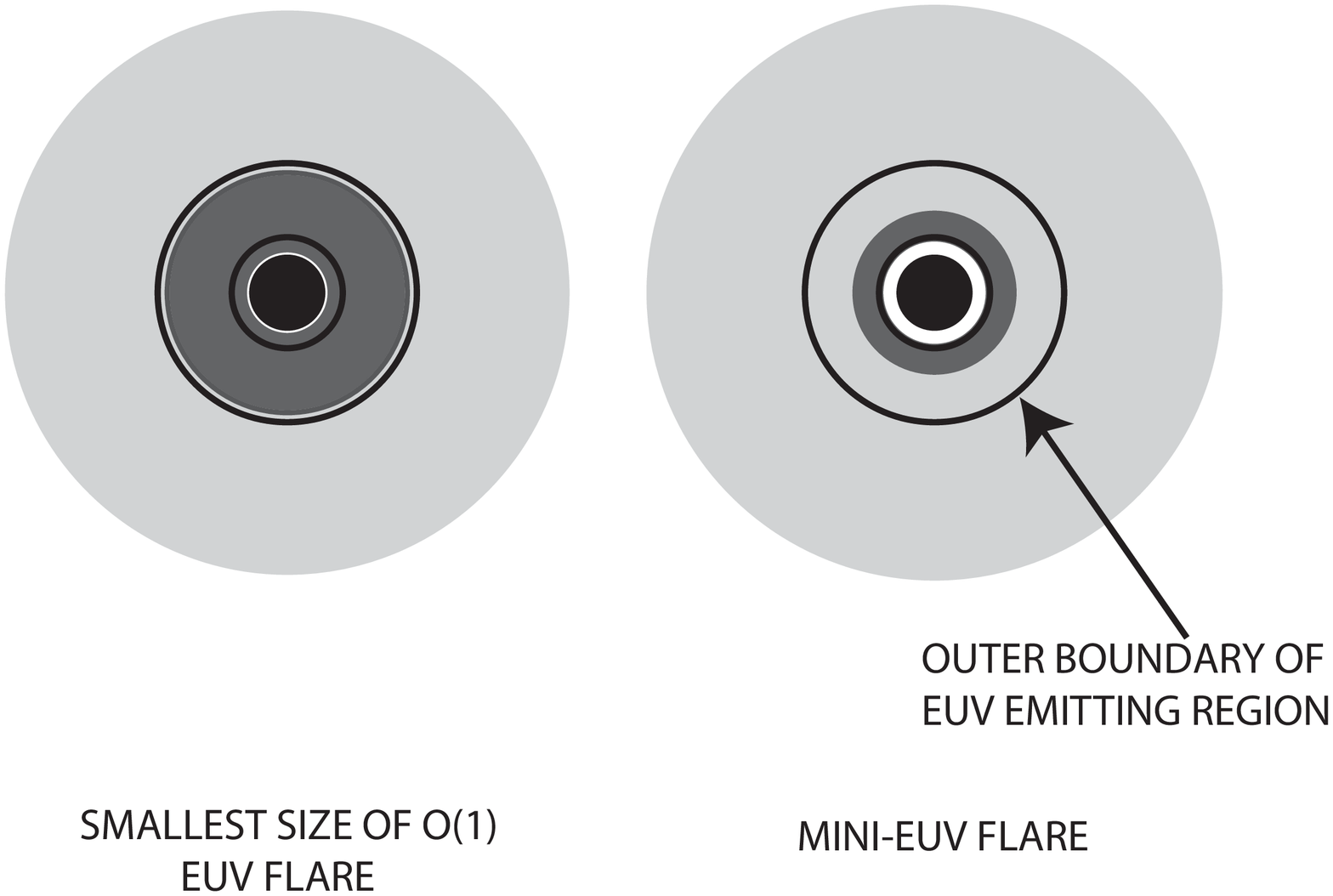}
\caption{The schematic shows a basic model in which the EUV
variability is primarily a consequence of the optically thin surface
layer of the inner region of a slim disk being evacuated and
replenished with new gas that has a different emissivity. The black
hole is the solid circle and the EUV emitting region is the annulus
between the inner edge of the disk and the demarcated circle. Darker
shading means higher emissivity of the gas.}
\end{center}
\end{figure}
\par The idea of associating the variability of the EUV continuum in luminous quasars to the emptying and refilling
of the inner annulus formed by the optically thin surface layer of
an accretion disk naturally produces dispersion in the amplitude of
variation and the time frame required for significant variation. In
principle, local physics or advection from larger radii can create
different magnitudes of the changes in the surface layer emissivity
(surface temperature) in the inner disk. Different time scales for
variability naturally arise from variations in black hole mass,
accretion rate and black hole spin. Based on Figures 6 - 8, the
crude model should be capable of explaining the coarse features of
EUV continuum variability in highly luminous quasars (Figures 2-5
and 9), the observed spread in variability magnitude and and the
observed spread in the time scale for significant variation.

\section{Discussion} This article explores the variability of the EUV
continuum emission in a sample of highly luminous quasars. The high
bolometric luminosity of these quasars was necessary in order to
insure high signal to noise absolute flux measurements of the EUV
continuum in HST and SDSS observations. The end result is not a
sample covering all of quasar parameter space, but only massive
black holes and high Eddington rates (most of the sample has $1.0
\times 10^{9} M_{\odot}<M_{bh} < 4.0 \times 10^{9} M_{\odot}$ and
$0.5< R_{\rm{Edd}} <1.1$). The analysis utilized pairs of spectral
observations from the SDSS and HST archives. Since the variability
that was detected is typically less than O(1), careful
considerations needed to be made in order to screen out false
variation that was a mere consequence of inadequate calibration
technique. The majority of this effort was consumed by the technical
details of the flux calibrations in order to minimize this
deleterious effect. In the end, we had a modest sample of 78 pairs
of quasar observations (54 SDSS pairs and 24 HST pairs). In spite of
the modest sample size, we are able to find a characteristic time
scale for measurable variability. At the crudest level, the
variability is larger when the sampling time between observations in
the quasar rest frame is $> 2\times 10^{7}$ sec compared to $<
1.5\times 10^{7}$ sec (at $>99.9\%$ level of statistical
significance according to a Kolmogorov-Smirnov test). Based on an
excess variance analysis, for time intervals $< 2\times 10^{7}$ sec
in the quasar rest frame, $10\%$ of the quasars show evidence of EUV
variability. Similarly, for time intervals $>2\times 10^{7}$ sec in
the quasar rest frame, $55\%$ of the quasars show evidence of EUV
variability (see Figure 5 and the related discussion). The amount of
variability levels off between $2.5\times 10^{7}$ sec and
$3.16\times 10^{7}$ sec (1 yr). This temporal behavior is indicative
of a threshold time scale for variability as opposed to a steady
monotonic increase. In Section 4, our findings were compared and
contrasted with similar behavior that is seen in photometry based
quasar structure functions. Marginal statistical evidence was
presented that the threshold time scale for significant variability
is spread out possibly as wide as $1.0\times 10^{7}$ sec to
$2.0\times 10^{7}$ sec.
\par This broad threshold was considered to be largely a consequence
of the spread of black hole mass and Eddington rates. We tried to
remove the black hole mass dependence by converting the time between
observations to geometrized units of black hole mass, M. In these
units the broad threshold appears to occur between 1800M and 2000M
(see the right hand frames of Figures 2 and 4). However, this did
not remove the dependence on Eddington rate. This could only be
achieved by appealing to theoretical models of accretion.

\par We pursued the possibility that this threshold time scale was directly associated with a characteristic spatial
dimension of the EUV emitting region in Section 5. This analysis
cannot be done purely empirically, but requires model dependent
assumptions. We chose the slim disk optically thick accretion model
as our fundamental assumption. In order to characterize the black
hole, we noted classical theoretical arguments and observational
analysis that for a large sample of highly luminous quasars most of
the central supermassive black holes should be spun up to near their
maximal allowed values \citep{bar70,elv02}. For these reasons, even
though we explored a broad range of black hole spins, our fiducial
value of spin was chosen to be $a/M=0.9$. In Figure 6, we showed
that $>75\%$ of the EUV emission is emitted from radii, $r < 9M$ and
$>50\%$ of the EUV emission is emitted from $r < 6M$ for $a/M=0.9$
with parameters that are characteristic of our sample, $1.0 \times
10^{9} M_{\odot}<M_{bh} < 4.0 \times 10^{9} M_{\odot}$ and $0.5<
R_{\rm{Edd}} <1.1$. Thus, the slim disk models, if they are correct,
imply that the EUV continuum variability is occurring as a
consequence of local physics in this region. Next, we tried to find
a plausible time scale within the slim disk models that is
associated with the dimensions of this region.

The most direct time scale to look for was the accreting gas infall
time to the plunge region. In order to represent the inflow velocity
of the optically thin surface layer responsible for the EUV
emission, we chose an extremely large viscosity parameter, $\alpha
=0.25$ in the optically thin surface layer that is responsible for
the EUV continuum (i.e roughly twice the value in the bulk of the
disk) as indicated in 3-D modeling of slim disks. The basic logic of
Section 5 can be shown in skeleton form. Our time interval is $T
\approx 2\times 10^{7}$ s or 1800M. The sample is dominated by
powerful quasars that have central black hole masses of
$M_{bh}\approx 2.5 \times 10^{9} M_{\odot}$ (see Tables 2 and 3)
which is $M\approx 3.5 \times 10^{14}$ cm in geometrized units. This
equates to a light travel time of $\approx 1.15 \times 10^{4}$ s.
This allows us to write the time interval in geometrized units as $T
\approx 1800$M. For radial inflow velocities in our fiducial model
with $a/M=0.9$ and $0.5< R_{\rm{Edd}} <1.1$ (based on Tables 2 and
3), the infall velocity is $V_{\rm{infall}}$ = 0.002c - 0.005c (see
the slim disk models in Figure 7). The infall distance in a time $T$
is $D=V_{\rm{infall}}T\approx 1.2 - 3.0 \times 10^{15}$ cm. Dividing
this by $M$ to get $D$ into geometrized units indicates that $D
\approx 3.5M - 8.5M$. We equated this distance to the radius from
which gas must fall inward to the black hole in order to produce the
observed variability time scale. For our fiducial value of the spin
$a/M \approx 0.9$, 75\% EUV emitting gas in the slim disk models is
located at $r < 9M$. Thus, this radial infall is a viable dynamic
for variability.
\par We tried two more analyses in order to see if the data was
consistent with the infall idea. The first was simply to plot the
amount of EUV flux density that emerges inside of a radius R versus
the infall time from that radius. This was plotted in the top frame
of Figure 8. We found that for black hole parameters that were
relevant to our sample, $T \approx 2\times 10^{7}$ s was associated
with infall from a radius of 5M - 6M. Furthermore, 50\% - 65\% of
the total EUV flux emerged from inside of this region. Thus, the
infall time scale was consistent with the slim disk model and the
observed variability. The most ambitious test was an attempt to
remove the affects of the Eddington ratio in Figure 9. We computed
the infall time as a function of radius in geometrized units based
on slim disk models associated with the estimated black hole mass
and Eddington rate for each quasar. We equated the infall time with
the time between observations. The radius from which the infall time
equals the time lag between observations was designated as the
infall radius. In Figure 9, we found that variability started to
increase significantly beyond infall radii $r>7.5 M$. This was the
third analysis (Figures 2-6, Figure 8 and Figure 9) that found
similar dimensions for the source of the EUV variability, indicating
consistency between the infall time from slim disk models and the
EUV continuum variability time scale.
\par Finally, we speculated that the agreement between the slim disk infall
time scale, the variability time scale and the location of the EUV
emission in slim disk models was not coincidental. We proposed a
very crude model in which the accretion of the optically thin
surface layer within $\gtrsim 5$M of the inner edge of the accretion
disk and replenishment with new gas of a different emissivity
(different surface temperature) was a viable candidate physical
mechanism for the EUV continuum variability in highly luminous
quasars.
\begin{acknowledgements}
We are indebted to Aleksander Sadowski for sharing his slim disk
results. We would like to thank Ian Evans for teaching us about the
limitations of FOS absolute flux measurements. We also benefited
from valuable information on COS flux calibrations from Todd Tripp
and Blair Savage. Todd Tripp also provided valuable insight into the
use of STIS spectra for flux measurements. We also were the
beneficiary of consult with Jack Brandt and David Soderblom on the
absolute flux calibration of HRS. BP notes that this research was
supported by ICRANet.
\end{acknowledgements}

\end{document}